\documentclass[aps,prx,10pt,floatfix,twocolumn,superscriptaddress,preprintnumbers,nofootinbib,notitlepage]{revtex4-2}
\usepackage{amsmath,amssymb,graphicx,xspace,bm}
\usepackage[normalem]{ulem}
\usepackage[utf8]{inputenc}
\usepackage[T1]{fontenc}
\usepackage{microtype}
\usepackage{siunitx}
\usepackage{dcolumn}
\usepackage{float}
\usepackage{xcolor}
\usepackage{hyperref}
\hypersetup{colorlinks=true,linkcolor=blue,citecolor=blue,urlcolor=blue}
\usepackage{mathtools}
\usepackage{orcidlink}

\preprint{}

\begin{document}

\rightline{FERMILAB-PUB-26-0492-T, CETUP2026-005}
\raggedbottom
\title{Astrophysical Neutrino Sources as Colliders}

\author{Carlos A. Arg\"uelles
\orcidlink{0000-0002-0192-8885}}
\email{carguelles@g.harvard.edu}
\affiliation{Harvard University, Department of Physics and Laboratory for Particle Physics and Cosmology, Cambridge, MA 02138, USA}

\author{P. S. Bhupal Dev
\orcidlink{0000-0003-4655-2866}}
\email{bdev@wustl.edu}
\affiliation{Department of Physics and McDonnell Center for the Space Sciences, Washington University, Saint Louis, MO 63130, USA}
\affiliation{PRISMA$^{++}$ Cluster of Excellence \& Mainz Institute for Theoretical Physics, 
Johannes Gutenberg-Universit\"{a}t Mainz, 55099 Mainz, Germany}
\author{Bhaskar Dutta \orcidlink{0000-0002-0192-8885}}
\email{dutta@tamu.edu}
\affiliation{
Mitchell Institute for Fundamental Physics and Astronomy,
Department of Physics and Astronomy, Texas A\&M University, College Station, TX 77843, USA
}

\author{Gonzalo Herrera \orcidlink{0000-0001-9250-8597}}
% \thanks{ORCID: \href{https://orcid.org/0000-0001-9250-8597}{0000-0001-9250-8597}}
\email{gonzaloh@mit.edu}
\affiliation{Kavli Institute for Astrophysics and Space Research, Massachusetts Institute of Technology, Cambridge, MA 02139, USA}
\affiliation{Harvard University, Department of Physics and Laboratory for Particle Physics and Cosmology, Cambridge, MA 02138, USA}

\author{Nicholas Kamp \orcidlink{0000-0001-9232-259X}}
\email{nkamp@g.harvard.edu}
\affiliation{Harvard University, Department of Physics and Laboratory for Particle Physics and Cosmology, Cambridge, MA 02138, USA}

\author{Jason Kumar \orcidlink{0009-0003-1230-2276}}
\email{jkumar@hawaii.edu}
\affiliation{Department of Physics and Astronomy, University of Hawai'i, Honolulu, HI 96822, USA}

\author{Stephan A. Meighen-Berger \orcidlink{0000-0001-6579-2000}}
\email{stephan-meighen-berger@uiowa.edu}
\affiliation{School of Physics, \href{https://ror.org/01ej9dk98}{The University of Melbourne}, Victoria 3010, Australia}
\affiliation{\href{https://ror.org/036jqmy94}{University of Iowa}, Iowa City, Iowa 52242, USA}

\author{Mudit Rai
\orcidlink{0000-0003-2876-809X}}
\email{muditrai@umich.edu}
\affiliation{
Mitchell Institute for Fundamental Physics and Astronomy,
Department of Physics and Astronomy, Texas A\&M University, College Station, TX 77843, USA
}
\affiliation{Leinweber Institute for Theoretical Physics, Department of Physics, University of Michigan, Ann Arbor, Michigan 48109, USA}

\author{Ian M. Shoemaker \orcidlink{0000-0001-5434-3744}}
\email{shoemaker@vt.edu}
\affiliation{Center for Neutrino Physics, Department of Physics, Virginia Tech, Blacksburg, VA 24061, USA}

\begin{abstract}
High-energy neutrinos arise from processes at large center-of-mass energies, offering a window to test physics at comparable scales or beyond those accessible in collider experiments on Earth. Here, we present a recipe for extracting two-sided bounds on the inelastic $pp$ and $p\gamma$ cross sections from neutrino point-source data, by independently constraining every astrophysical input (cosmic-ray luminosities and target densities) through electromagnetic observations or theoretical arguments.  
The cross section is then the only remaining free parameter. Applying this framework to the IceCube associations with TXS~0506+056, NGC~1068, and the Galactic Plane, to a stacked population of eleven X-ray bright Seyfert galaxies, to the ultra-high-energy KM3NeT event KM3-230213A, and to projected observations of ultra-high-energy neutrinos, we obtain constraints that span center-of-mass energies from $\sqrt{s}\sim 1$~GeV to $\sim 10^{5}$~GeV, some of which are well beyond the reach of the LHC and, for the $p\gamma$ channel, beyond HERA.
Several of these bounds are more stringent than unitarity limits.
\end{abstract}

\maketitle

%=====================================================================
\section{Introduction}
%=====================================================================

The identification of individual neutrino sources and the observation of ultra-high-energy neutrinos, notably, TXS~0506+056~\cite{IceCube:2018cha}, NGC~1068~\cite{IceCube:2022der} and a population of X-ray bright Seyfert galaxies~\cite{IceCube:2024dou} by IceCube, and the $\sim$220~PeV event KM3-230213A at KM3NeT~\cite{KM3NeT:2025npi,KM3NeT:2025ccp}, has turned neutrino telescopes into probes~\cite{Arguelles:2024ncf} of hadronic interactions at center-of-mass energies that overlap with, and in several cases exceed, those accessible at terrestrial colliders. For instance, a 290~TeV neutrino from TXS~0506+056 produced in $pp$ collisions implies a center-of-mass energy $\sqrt{s}\simeq 3.3$~TeV, comparable to the LHC; the 220 PeV KM3NeT event would probe $\sqrt{s}\simeq 91$~TeV, far beyond $\sqrt{s}_{\rm LHC}=14$~TeV.

Recent literature exploits these extreme environments to search for beyond-the-Standard-Model (BSM) physics such as dark matter--cosmic-ray (CR) interactions~\cite{Herrera:2023nww, Gustafson:2024aom,
DeMarchi:2024riu,  Mishra:2025juk, Kantzas:2025huu, Hussein:2025llu, Meighen-Berger:2025hrq}, CR boosted dark matter~\cite{Wang:2021jic, DeMarchi:2025uoo, Gustafson:2025dff}, dark matter--neutrino scattering~\cite{Arguelles:2017atb,IceCube:2022clp,Cline:2022qld,Ferrer:2022kei,  Fujiwara:2023lsv, 
Fujiwara:2024qos,Zapata:2025huq, Pompa:2025lbf,  Tseng:2024akh, Bertolez-Martinez:2025trs,Mondol:2025uuw}, dark matter--photon signals~\cite{Ferrer:2022kei, Herrera:2025gpm}, dark-matter production~\cite{Dev:2025czz}, quasi-Dirac neutrinos and ultra-long-baseline neutrino oscillations~\cite{Rink:2022nvw,Carloni:2022cqz,Dixit:2024ldv,Carloni:2025dhv, MacDonald:2025jbm}, and graviton/black hole production~\cite{Ettengruber:2025kat}. 
Several studies have also used these high-energy neutrino events to constrain the neutrino--nucleon cross section at the detector~\cite{Alvarez-Muniz:2001efi, IceCube:2017roe, 
Bustamante:2017xuy, IceCube:2020rnc, Bai:2025pef, Bertolez-Martinez:2026bzj, Palmisano:2026sid}.
These bounds, however, underscore the importance of disentangling the degeneracies in the neutrino flux arriving at Earth from the neutrino-nucleon cross section, and do not test the same channels that colliders predominantly do.
Rather than measuring the neutrino cross section at the detector, we bound the hadronic production cross sections---$pp$ and $p\gamma$---at the source. 
The key observation is that all astrophysical inputs entering the neutrino-luminosity equation can be independently constrained: the CR luminosity is bounded from above by the Eddington limit (or its relativistic generalization for blazars) and from below by the minimum power required to sustain the observed neutrino and X-ray fluxes; the target column density is fixed by X-ray spectroscopy, bolometric measurements, Thomson-depth measurements, or variability arguments. 
With these inputs pinned down by electromagnetic data and theoretical considerations, the hadronic cross section remains as the only free parameter, and neutrino observations bound it from both sides.  
This provides a recipe that any future neutrino point-source detection can follow to test QCD, complementary to colliders. 
Importantly, unlike colliders, neutrino sources operate across several orders of magnitude in center-of-mass energies, providing bounds at values that have not been scanned by colliders, which operate at fixed center-of-mass energies.

The bounds we derive are limited by the small number of identified neutrino sources and by the systematics of their inferred astrophysical properties---CR luminosities, target column densities, and source geometries.
All of the observations considered here, however, share a single underlying ingredient: the same inelastic $pp$ or $p\gamma$ cross section.
As more sources are identified and as electromagnetic and multi-messenger observations sharpen the determination of their astrophysical inputs, the degeneracies between source modeling and particle physics will progressively be disentangled, and the cross-section determination will tighten correspondingly.
In this paper we make the distinct degeneracies explicit on a source-by-source basis, derive the first two-sided upper and lower limits on the inelastic $pp$ and $p\gamma$ cross sections from neutrino point-source observations, and systematically quantify the center-of-mass energies probed by each source.
Our work thus serves as a roadmap for the measurement of the $pp$ and $p\gamma$ cross sections at energies complementary to, and in several cases beyond, the reach of terrestrial colliders.

The remainder of this paper is organized as follows.
In Sec.~\ref{sec:pp_bounds} and Sec.~\ref{sec:pgamma_bounds} we derive the upper and lower bounds on $\sigma_{pp}$ and $\sigma_{p\gamma}$, respectively, and apply them to TXS~0506+056, NGC~1068, and the Galactic plane.
Section~\ref{sec:seyfert} extends the analysis to the stacked Seyfert population recently reported by IceCube.
Section~\ref{sec:projected} presents projected bounds from the KM3NeT event KM3-230213A and from future ultra-high-energy neutrino observations.
Section~\ref{sec:laboratory} compares our bounds with laboratory and astrophysical data and the Froissart--Martin unitarity limit.
We conclude in Sec.~\ref{sec:conclusions}. Appendix~\ref{app:pythia} describes our neutrino energy fraction calculation using Pythia; Appendix~\ref{app:spectral} discusses the dependence of the center-of-mass energy on the CR spectral index and neutrino spectral width;  Apendix~\ref{app:crcr} considers the case of CR--CR collisions.

%=====================================================================
\section{Bounds on the $pp$ cross section}
\label{sec:pp_bounds}
%=====================================================================
If the observed neutrino flux from a source is produced predominantly via inelastic $pp$ collisions producing charged pions, the all-flavor neutrino luminosity is related to the injected CR  power by
\begin{equation}
L_\nu \simeq \left( \frac{3\,\xi_\nu^{pp}}{\kappa_{pp}} \right) \, f_{pp} \, L_{\rm CR}\,, 
\label{eq:Lnu_pp}
\end{equation}
where $\xi_\nu^{pp} \simeq 0.086-0.099$ (See our Pythia simulations in Appendix \ref{app:pythia}) is the fraction of the proton energy carried by neutrinos of a single flavor, summed over the full multiplicity of neutrinos produced per collision (set by pion decay kinematics and independent of the $pp$ cross section), the factor of 3 accounts for three neutrino flavors after oscillation averaging, and $L_{\rm CR}$ is the proton injection luminosity.  
The inelasticity $\kappa_{pp} \simeq 0.5$~\cite{Kelner:2006tc,Murase:2013rfa} is the average fraction of the proton energy 
lost per inelastic $pp$ scatter and  $f_{pp}$ is the overall fraction of the proton's injection energy lost through 
inelastic $pp$ scatters. 
%The parameter $\xi_\nu$ is the fraction of the incoming proton energy carried away by each neutrino flavor per collision. 

Leading-pion kinematics give $\xi_\nu^{pp} \simeq 0.03$--$0.05$~\cite{Kelner:2006tc}; we run dedicated \textsc{Pythia\,8}~\cite{Skands:2014pea} simulations, which capture the full multiplicity spectrum and kaon contributions, finding $\langle\xi_\nu^{pp} \rangle(\sqrt{s}) \simeq 0.086$--$0.099$ (see Appendix~\ref{app:pythia} and Figs.~\ref{fig:pythia_dist} and \ref{fig:pythia_props}), and we adopt these values throughout.
In general, a proton traversing the source loses energy to multiple channels $i$: inelastic $pp$, elastic $pp$, and $p\gamma$ interactions. 
However, in the energy range which we consider, elastic scattering is negligible.
The total fractional energy loss is $f_{\rm tot} = 1 - \exp\!\bigl(-\sum_i \kappa_i~\sigma_i\, N_i^{\rm eff}\bigr)$, and the fractional energy loss 
to channel $i$ can be expressed as 
\begin{equation}
f_i =\frac{\kappa_i~\sigma_i\, N_i^{\rm eff}}{\sum_j \kappa_j~\sigma_j\, N_j^{\rm eff}} f_{tot} .
\end{equation}

In the optically thin regime, $\sum_i \sigma_i\,N_i^{\rm eff} \ll 1$, the channels decouple and each contributes independently: 
$f_i \simeq \kappa_i~\sigma_i\,N_i^{\rm eff}$. 
%Since elastic scattering produces no pions and therefore no neutrinos, only the inelastic channel contributes to the neutrino yield. 
We thus isolate
\begin{equation}
%f_{pp}(E_p) \simeq 1 - \exp\left(-\sigma_{pp} \, N_p^{\rm eff}\right),
f_{pp}(E_p) \simeq \kappa_{pp}~\sigma_{pp} \, N_p^{\rm eff}\  ,
\label{eq:fpp}
\end{equation}
where $\sigma_{pp}$ is the inelastic $pp$ cross section and $N_p^{\rm eff}$ is the effective gas column density traversed by the accelerated protons in the neutrino production region.

We then find that the neutrino flux is linearly proportional to the cross section.

In the optically thick regime ( $\sum_i \sigma_i\,N_i^{\rm eff} \gg 1$), however, we would find that 
$f_{tot} \rightarrow 1$.  In this limit, the connection between the neutrino flux and the proton-proton and proton-photon 
scattering cross sections is much more tenuous.  Essentially, all of the proton's energy is lost to inelastic scattering, 
and the scattering cross sections only determine the length scale over which the energy is lost.  More precisely,  
summing over all channels, we would find $L_\nu \sim \overline{\xi_\nu / \kappa}~ L_\mathrm{CR}$, 
where $\overline{\xi_\nu / \kappa} \sim 0.2$ is the weighted average of $\xi_\nu^i / \kappa_i$ over all channels.  The only leverage one 
would have in determining the cross section is the precise weighting of this average.  However, a target cannot be in the 
optically thick limit if $L_\nu \lesssim 0.2~L_\mathrm{CR, min}$, where $L_\mathrm{CR, min}$ is the minimum possible 
CR luminosity.  This criterion will be easily satisfied for the targets we consider,\footnote{The 
only exception is the flaring event from the source 
TXS~0506+056.  For this source, we will assume that we are in the optically-thin limit for the purpose of obtaining an upper bound on the 
cross section.  In any case, this upper bound will not be too far from the unitarity bound, so the cross section could not be 
much larger, even in the optically-thick limit.} so we will be justified 
in simply assuming that we are in the optically-thin limit.

More generally, if we take $\sigma_{pp}$ to be the same as the Standard-Model (SM) inelastic $pp$ cross section 
($\sigma_{pp}^\mathrm{SM}$), then  for each source considered here, the assumed column densities satisfy 
$\sigma_{pp}^{\rm SM}\,N_p^{\rm eff}\lesssim 1$:

for TXS~0506+056 (blazar jet) $\sim 10^{-3}$--$10^{-2}$~\cite{Keivani:2018rnh}; for NGC~1068 (Compton-thick corona) $\sim 0.06$--$0.4$, consistent with partial but not full calorimetry~\cite{Murase:2022dog,Inoue:2019yfs}; for the Galactic plane $\sim 10^{-2}$--$10^{-1}$~\cite{AMS:2016brs}; and for the Seyfert stack $\sim 0.06$--$0.2$~\cite{Fabian:2015MNRAS,Murase:2019vdl}. 

\subsection{Band width and dominant uncertainties}

To produce the observed neutrino luminosity in the optically-thin limit, the proton cooling efficiency must 
satisfy $L_\nu \kappa_{pp} / (3\,\xi_\nu^{pp} \, L_{\rm CR,min}) \gtrsim f_{pp} \gtrsim L_\nu \kappa_{pp} / (3\,\xi_\nu^{pp} \, L_{\rm CR,max})$, 
where $L_{\rm CR,max(min)}$ denotes the maximum (minimum) CR power allowed by the energetics of the source.
Inverting this condition, we obtain ,
\begin{equation}
\sigma_{\rm hi/lo} \simeq \frac{L_\nu}{3\,\xi_\nu^{pp}  \, N_{p,\,\min/\max}^{\rm eff} \, L_{\rm CR,\,min/max}},
\label{eq:sigma_pp}
\end{equation}
where $N_{p,\min}^{\rm eff}$ and $N_{p,\max}^{\rm eff}$ denote the minimum and maximum of the allowed range for $N_p^{\rm eff}$.
We thus have a band width
\begin{equation}
\frac{\sigma_{\rm hi}}{\sigma_{\rm lo}} \approx
\underbrace{\frac{L_{\rm CR,max}}{L_{\rm CR,min}}}_{\text{CR luminosity}} \;\times\;
\underbrace{\frac{N_{p,\max}^{\rm eff}}{N_{p,\min}^{\rm eff}}}_{\text{column density}}.
\label{eq:bandwidth}
\end{equation}
The two principal astrophysical uncertainties---the CR power budget and the effective target column density---thus multiply to set the dynamic range of the allowed band. For the Seyfert stack, $L_{\rm CR}$ spans a factor of~10 (from the CR-to-intrinsic-X-ray luminosity ratio $\eta_{\rm CR} \equiv L_{\rm CR}/L_X^{\rm intr} \in [1,10]$) and $N_p^{\rm eff}$ spans a factor of~3 (from the coronal Thomson optical depth $\tau_T \in [1,3]$), yielding a band width of~$\sim 30$. For the Galactic plane, $L_{\rm CR}$ spans roughly half a decade and $N_p^{\rm eff}$ roughly half a decade, also yielding~$\sim 30$. Narrowing either uncertainty directly tightens the cross-section constraints.

\subsection{Center-of-mass energy}

The center-of-mass energy probed at each source is determined by the observed neutrino energy range $[E_{\nu,\mathrm{lo}},\, E_{\nu,\mathrm{hi}}]$, taken from IceCube measurements or projected sensitivities (Table~\ref{tab:sources_transposed}), and the neutrino production kinematics.  For $pp$ collisions with a target proton at rest, the proton energy in the lab frame, $E_p$, is related to the neutrino energy by $E_p = E_\nu / \xi_\nu^{pp} $, and
\begin{equation}
\sqrt{s} = \sqrt{2\,m_p\,E_p + 2\,m_p^2}\,,
\label{eq:sqrts}
\end{equation}
so $\sqrt{s}$ depends on $\xi_\nu^{pp}$, the mean fraction of proton energy carried by a neutrino of a given flavor.  Typical values commonly adopted in the literature span the range $\xi_\nu^{pp} \simeq 0.03$--$0.05$ per flavor from leading-pion decay kinematics~\cite{Kelner:2006tc}.  Here we perform dedicated Monte Carlo simulations with \textsc{Pythia\,8}~\cite{Sjostrand:2014zea}, finding an energy-dependent total per-flavor fraction $\langle\xi_\nu^{pp} \rangle(\sqrt{s}) \simeq 0.086$--$0.099$ (see Appendix and Figs.~\ref{fig:pythia_dist},~\ref{fig:pythia_props}), roughly a factor of~$\sim 1.8$ larger than the analytical estimate.  The difference arises because the Monte Carlo sums over the full neutrino multiplicity per event---including sub-leading pions that individually carry less energy but collectively boost the total per-flavor yield---as well as kaon and charmed-meson contributions. We note that $\xi_\nu$ as defined is the \emph{total} energy fraction carried by all neutrinos of a single flavor, summed over the multiplicity; it is not the energy of an individual detected neutrino. The mapping $E_p = E_\nu/\xi_\nu$ is therefore approximate, since a single detected neutrino carries less energy than the per-flavor total. This approximation is conservative: it slightly underestimates $E_p$ and thus $\sqrt{s}$, meaning the true center-of-mass energies probed are higher than shown. The cross-section bounds themselves depend on $\xi_\nu$ only through the energy budget [cf.~Eq.~\eqref{eq:Lnu_pp}] and are unaffected by this mapping.

We adopt the \textsc{Pythia}-derived $\langle\xi_\nu\rangle(\sqrt{s})$ throughout this work.  For each source, the two endpoints of the observed neutrino energy range are mapped to $\sqrt{s}$ values by iterating $E_p = E_\nu / \langle\xi_\nu\rangle(\sqrt{s})$, with $\sqrt s$ given by Eq.~\eqref{eq:sqrts},  
until convergence, yielding a self-consistent $\sqrt{s}$ band.  This ``mean-$\xi$'' prescription assigns a single characteristic $\sqrt{s}$ to each observed $E_\nu$; in reality, the neutrino energy distribution $E_\nu\,dN/dE_\nu$ at a given $\sqrt{s}$ has finite width (Fig.~\ref{fig:pythia_dist}b), so an observed $E_\nu$ receives contributions from a \emph{range} of center-of-mass energies.  We quantify this broadening in Appendix~\ref{app:spectral} and Fig.~\ref{fig:spectral_index}b by computing the $\sqrt{s}$ range corresponding to neutrinos within a given fraction of the spectral peak ($E_\nu\,dN/dE_\nu \geq f \times \mathrm{peak}$, for $f = 1\%$, $10\%$, $60\%$, $90\%$).  For a $10\%$ threshold, the effective $\sqrt{s}$ band broadens by a factor of $\sim 2$--$3$ relative to the mean-$\xi$ value; the cross-section bounds themselves are essentially independent of this broadening, since they depend on $\xi_\nu$ which varies only weakly over the relevant $\sqrt{s}$ range.

An additional source of uncertainty in $\sqrt{s}$ arises from the spectral index $\Gamma$ of the CR injection spectrum $dN_{\mathrm{CR}}/dE \propto E^{-\Gamma}$, which weights the luminosity-averaged neutrino energy within the observed band.  This effect is discussed in Appendix~\ref{app:spectral} and Fig.~\ref{fig:spectral_index}a.

\begin{figure*}[t!]
\centering
\includegraphics[width=0.8\textwidth]{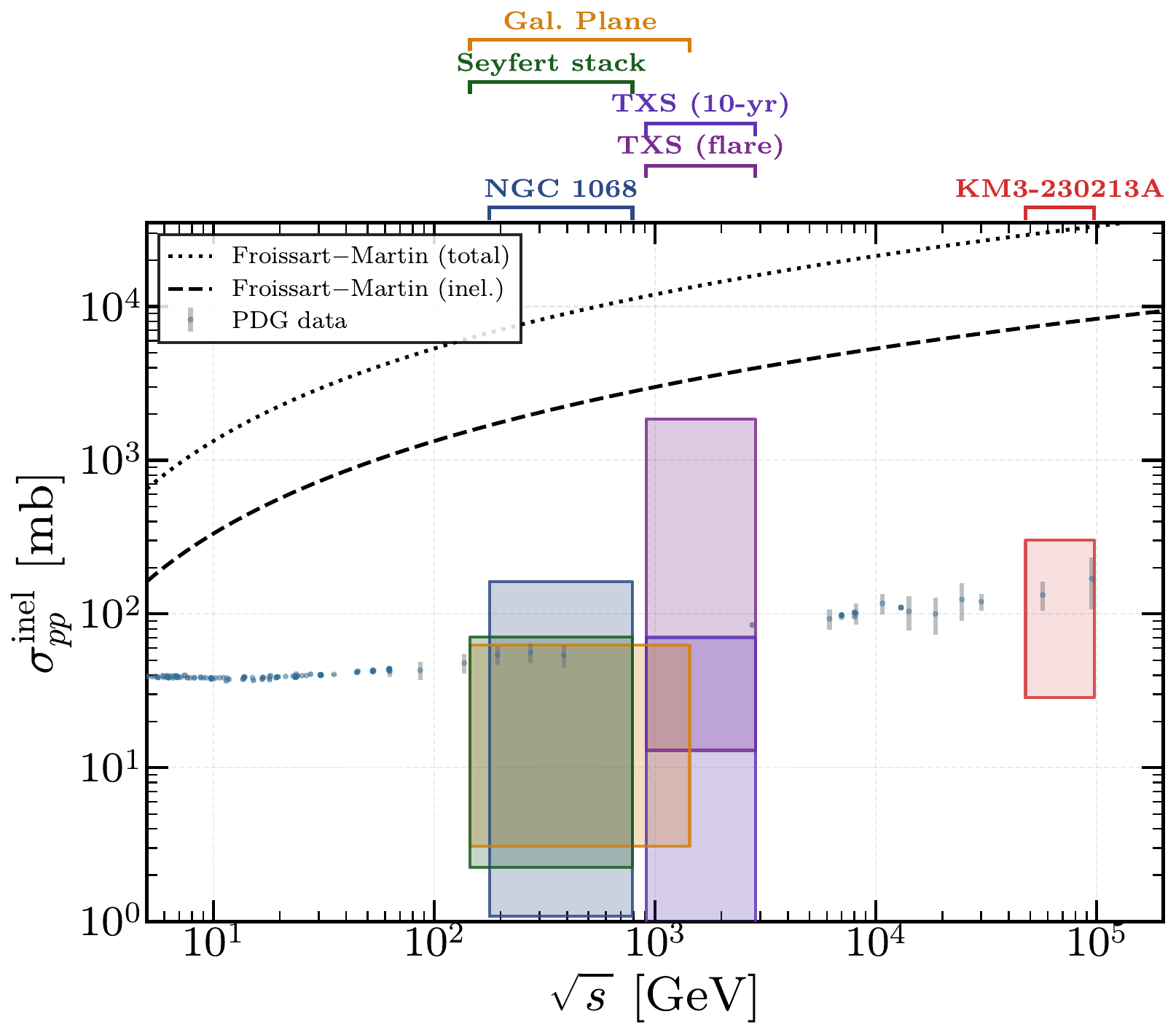}
\caption{\textbf{\textit{Allowed values of the inelastic $pp$ cross section from neutrino observations.}} The blue band shows the constraint from NGC~1068. The purple band shows the TXS~0506+056 constraint from the 2014--15 flare luminosity, and the indigo band shows the constraint using the 10-year time-integrated luminosity, which is a factor $\sim 25$ lower and yields a correspondingly tighter lower bound; both assume a BZ+MAD beaming-corrected $L_{\rm CR,max}^{\rm iso} \simeq 3.4 \times 10^{49}$~erg\,s$^{-1}$ (see text). The orange band shows the Galactic Center Plane constraint. The green band shows the constraint from the stacked Seyfert population (excluding NGC~1068). The red band shows the KM3-230213A projection adopting a ``standard-candle'' idealization: target column and CR luminosity fixed to their best-estimate values, band width driven only by the neutrino-luminosity uncertainty, centred on the SM prediction at $\sqrt{s} \sim 50$--$100$~TeV. For comparison, we show PDG data from collider and fixed-target experiments~\cite{ParticleDataGroup:2024cfk} and the Froissart--Martin unitarity bound~\cite{Martin:2009pt} for the total (dotted) and inelastic (dashed) cross sections. Center-of-mass energies are computed self-consistently using the \textsc{Pythia}-derived $\langle\xi_\nu\rangle(\sqrt{s})$.}
\label{fig:limits_pp}
\end{figure*}
\begin{figure*}[t!]
\centering
\includegraphics[width=0.8\textwidth]{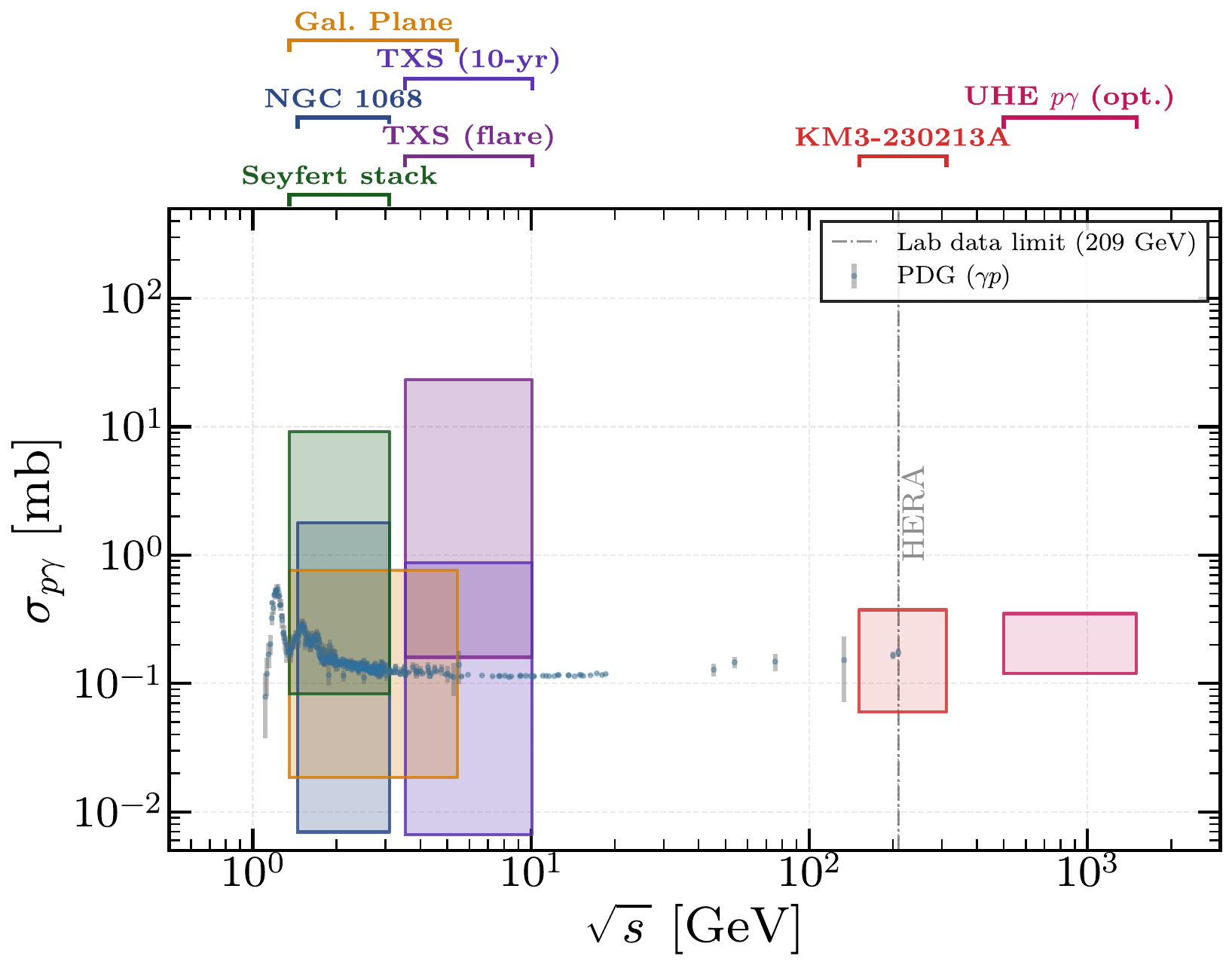}
\caption{\textbf{\textit{Allowed values of the inelastic $p\gamma$ cross section from neutrino observations, assuming a 10~keV X-ray target photon field.}} The target photon field density is computed from the absorption-corrected intrinsic X-ray luminosity of each source (from the BASS catalog~\cite{Koss:2022ApJS}), rather than from the directly observed flux; the absorption correction accounts for photoelectric attenuation by cold line-of-sight gas. We further assume that the coronal region is transparent to $\gamma\gamma$ pair production at the TeV energies relevant for the associated neutrinos, so that the inferred photon-field density is a good proxy for the target traversed by accelerated protons. Source colors and TXS~0506+056 flare/10-year conventions are as in Fig.~\ref{fig:limits_pp}. The red and pink bands (KM3-230213A and UHE $p\gamma$ projections) adopt a ``standard-candle'' idealization: the target photon column and CR luminosity are fixed to their best-estimate values and only the neutrino luminosity uncertainty is propagated, yielding narrow bands centered on the SM prediction at the corresponding $\sqrt{s}$. The UHE band specifically shows the most optimistic configuration: perfect handle on $L_\nu$ and $N_\gamma^{\rm eff}$ and highest attainable energy ($\sqrt{s} \sim 0.5$--$1.5$~TeV), beyond the HERA reach. For comparison, we show photoproduction data from HERA and fixed-target experiments~\cite{ParticleDataGroup:2024cfk}.}
\label{fig:limits_pgamma}
\end{figure*}

\subsection{Application to individual sources}

\paragraph*{TXS~0506+056.---}
TXS~0506+056 is a BL~Lac blazar ($M_{\rm BH} \simeq 3 \times 10^8\,M_\odot$, $L_{\rm Edd} \simeq 3.8 \times 10^{46}$~erg\,s$^{-1}$) whose relativistic jet is closely aligned with our line of sight. The observed neutrino flux during the 2014--2015 flare, $E_\nu \simeq 40$--$400$~TeV, implies an isotropic-equivalent neutrino luminosity $L_\nu^{\rm flare} \simeq 1.2 \times 10^{47}$~erg\,s$^{-1}$~\cite{IceCube:2018cha}, which exceeds $L_{\rm Edd}$ by a factor $\sim 3$. For blazars, however, the isotropic-equivalent luminosity can far exceed $L_{\rm Edd}$ without violating any physical limit: the jet emission is relativistically beamed into a cone of opening angle $\sim 1/\Gamma_{\rm jet}$, so an observer on-axis infers $L_{\rm iso} \sim 2\Gamma_{\rm jet}^2\,L_{\rm true}$ for a continuous jet~\cite{Urry:1995mg}. The true jet power is bounded by the Blandford--Znajek (BZ) mechanism~\cite{Blandford:1977ds}, which extracts rotational energy from the black hole; in the magnetically arrested disk (MAD) state the jet efficiency can reach $\eta_{\rm jet} \sim 100$--$300\%$ of $\dot{M}c^2$ for maximal spin~\cite{Tchekhovskoy:2011zx}, so $L_{\rm jet} \lesssim \eta_{\rm jet} \, L_{\rm Edd}$. Combined with relativistic beaming ($\Gamma_{\rm jet} \sim 10$--$20$ for BL~Lac jets~\cite{Ghisellini:2014pwa}), the maximum conceivable isotropic-equivalent CR luminosity is
\begin{equation}
L_{\rm CR,max}^{\rm iso} \sim 2\Gamma_{\rm jet}^2 \, \eta_{\rm jet} \, L_{\rm Edd}\,.
\end{equation}
For our fiducial values $\Gamma_{\rm jet} = 15$ and $\eta_{\rm jet} = 2$, this gives $L_{\rm CR,max}^{\rm iso} \simeq 3.4 \times 10^{49}$~erg\,s$^{-1}$ ($\simeq 900\,L_{\rm Edd}$), consistent with multi-messenger models of TXS~0506+056 that adopt isotropic-equivalent proton luminosities $L_p^{\rm iso} \sim 10^{49}$--$10^{50}$~erg\,s$^{-1}$~\cite{Keivani:2018rnh,Cerruti:2018tmc}, corresponding to sub-Eddington true jet powers after the beaming factor is divided out (see also~\cite{Gustafson:2025dff,Dev:2025czz}). The minimum cosmic-ray luminosity is set by energy conservation: sustaining the observed neutrino output requires $L_{\rm CR,min} \simeq 2.6\times 10^{48}$~erg\,s$^{-1}$.

With this beaming-corrected ceiling, the ratio $L_\nu^{\rm flare}/(3\,\xi_\nu^{pp} \, L_{\rm CR,max}^{\rm iso}) \simeq 0.013 \ll 1$ (using $\xi_\nu^{pp} \simeq 0.09$ from \textsc{Pythia}), so a finite two-sided bound on $\sigma_{pp}$ exists even for the 2014--15 flare epoch. The resulting band spans $\sigma_{pp} \simeq 13$--$1800$~mb at $\sqrt{s} \simeq 0.9$--$2.8$~TeV (Fig.~\ref{fig:limits_pp}). The allowed band is broad; spanning roughly two orders of magnitude, because
the enormous $L_{\rm CR,max}^{\rm iso}$ available to blazars makes it easy to
produce the observed neutrino flux even with a modest cross section, placing the
lower bound near the SM value. Averaged over the full $\sim$10-year IceCube dataset, the time-integrated neutrino luminosity is $L_\nu^{\rm 10yr} \simeq 5 \times 10^{45}$~erg\,s$^{-1}$~\cite{IceCube:2022der}, roughly a factor of $25$ below the flare value, yielding a tighter two-sided bound $\sigma_{pp} \simeq 0.5$--$70$~mb at the same $\sqrt{s}$ range.

TXS~0506+056 has no evidence for significant intrinsic X-ray absorption~\cite{Keivani:2018rnh}. The effective proton column density is not directly measured, but can be constrained using variability and multi-wavelength transparency arguments. The duration of the enhanced activity, $t_{\rm var} \sim \mathcal{O}(10^2)$~days~\cite{IceCube:2018cha}, constrains the characteristic size of the interaction region by causality to $\ell \lesssim c \, t_{\rm var} \, \delta$, where $\delta \sim 10$ is the Doppler factor of the jet, implying $\ell \lesssim (1$--$3) \times 10^{18}$~cm. Requiring that the baryonic material within this region not overproduce thermal emission constrains $n_p \lesssim 10^{6\text{--}7}$~cm$^{-3}$~\cite{Murase:2018iyl, Celotti:2007rb}, giving $N_p^{\rm eff} \lesssim 10^{23}$--$10^{24}$~cm$^{-2}$. We adopt this range as representative; since only an upper bound on 
%$n_p$ 
$N_p^{\rm eff}$ is available, the upper constraint on $\sigma_{pp}$ for this source should be interpreted with caution.

\paragraph*{NGC~1068.---}
A steady neutrino flux was observed in the energy range $E_\nu \simeq 1.5$--$30$~TeV~\cite{IceCube:2022der}. The CR luminosity is bracketed by the intrinsic X-ray luminosity ($\sim 5 \times 10^{43}$~erg\,s$^{-1}$) from below and a fraction of the bolometric luminosity ($L_{\rm bol} \simeq 10^{45}$~erg\,s$^{-1}$) from above~\cite{Herrera:2023nww, Murase:2019vdl, Murase:2022dog}. Our choice is conservative, as AGN corona models that successfully explain the IceCube signal require $L_{\rm CR}$ up to a few~$\times 10^{44}$~erg\,s$^{-1}$~\cite{Murase:2022dog, Inoue:2019yfs}. We adopt $L_{\rm CR} \in [5 \times 10^{43},\,  10^{45}]$~erg\,s$^{-1}$.
NGC~1068 is a well-established Compton-thick AGN, with line-of-sight hydrogen column densities $N_H \sim 10^{25}~\mathrm{cm}^{-2}$ from NuSTAR, XMM-Newton, and Chandra spectroscopy~\cite{Marinucci:2015fqo, Bauer:2014rla}. The effective proton column density depends on whether CRs interact primarily within the corona ($\tau_T \sim 1$, $N_p^{\rm eff} \sim 1.5 \times 10^{24}$~cm$^{-2}$) or also traverse the Compton-thick material. We adopt
\begin{equation}
N_{p,\,\mathrm{NGC\,1068}}^{\rm eff} \sim 1.5 \times 10^{24}\text{--}10^{25}~\mathrm{cm}^{-2},
\end{equation}
spanning from the coronal Thomson depth to the full Compton-thick column~\cite{Murase:2022dog, Inoue:2019yfs, Marinucci:2015fqo}. With these parameters, the SM cross section lies within the allowed band.

\paragraph*{Galactic plane.---}
IceCube has reported evidence for diffuse neutrino emission from the Galactic plane at $E_\nu \simeq 1$--$100$~TeV~\cite{IceCube:2023ame}, consistent with hadronic interactions of Galactic CRs with interstellar gas. The Galactic-plane neutrino luminosity $L_\nu \simeq 5 \times 10^{38}$~erg\,s$^{-1}$ (all-flavor, model-independent)~\cite{IceCube:2023ame} is measured directly by IceCube. The CR luminosity of the Milky Way is independently bounded by supernova energetics: with $\sim 2$--$3$ core-collapse supernovae per century each injecting $\sim 10^{51}$~erg at $\sim 10\%$ efficiency into CRs~\cite{Strong:2010pr}, one obtains $L_{\rm CR} \simeq 5 \times 10^{40}$--$2 \times 10^{41}$~erg\,s$^{-1}$. The effective proton column density traversed by CRs is constrained by the boron-to-carbon (B/C) grammage measurements at TeV energies to $N_p^{\rm eff} \simeq 6 \times 10^{23}$--$3 \times 10^{24}$~cm$^{-2}$~\cite{AMS:2016brs, CALET:2022vro}. All three inputs---$L_\nu$, $L_{\rm CR}$, and $N_p^{\rm eff}$---are therefore determined independently of $\sigma_{pp}$. The resulting bounds on $\sigma_{pp}$ cover $\sqrt{s} \sim 150$--$1400$~GeV.

\medskip
Together, these relations define a band of allowed values for $\sigma_{pp}$ for each neutrino source. The input parameters and assumptions for each source are summarized in Table~\ref{tab:sources_transposed}. For TXS~0506+056, the BZ+MAD beaming-corrected CR luminosity ceiling yields finite two-sided bounds for both the 2014--15 flare and the 10-year time-integrated luminosity; the latter provides the tighter constraint. NGC~1068 and the Galactic plane provide two-sided bands at $\sqrt{s} \sim 150$--$1400$~GeV, with the SM cross section lying within the allowed region. Results are shown in Fig.~\ref{fig:limits_pp}.

%=====================================================================
\section{Bounds on the $p\gamma$ cross section}
\label{sec:pgamma_bounds}
%=====================================================================

Neutrino production via inelastic photohadronic interactions, $p\gamma \to \pi X$, may dominate in sources with intense radiation fields, such as AGN coronae and blazar jets. In close analogy to the hadronuclear case [cf.~Eq.~\eqref{eq:Lnu_pp}], the neutrino luminosity is
\begin{equation}
L_\nu \simeq \left( \frac{3\,\xi_\nu^{p\gamma} }{\kappa_{p\gamma}} \right)\, f_{p\gamma} \, L_{\rm CR},
\label{eq:Lnu_pgamma}
\end{equation}
where $\xi_\nu^{p\gamma}$ is the fraction of the proton energy transferred to neutrinos of a single flavor through photomeson production and pion decays, the factor of~3 accounts for three neutrino flavors after oscillations, 
$\kappa_{p\gamma} \simeq 0.2$ is the inelasticity~\cite{Mucke:1998mk}, 
and $f_{p\gamma}$ is the effective fractional energy loss due to $p\gamma$ interactions,  
which in the optically-thin limit is given by
\begin{equation}
%f_{p\gamma} \simeq 1 - \exp\left(-\sigma_{p\gamma} \, N_\gamma^{\rm eff}\right),
f_{p\gamma} \simeq \kappa_{p\gamma}~\sigma_{p\gamma} \, N_\gamma^{\rm eff} ,
\label{eq:fpgamma}
\end{equation}
where $\sigma_{p\gamma}$ is the inelastic $p\gamma$ cross section, $N_\gamma^{\rm eff}$ is the effective photon column density in the interaction region, and $\xi_\nu^{p\gamma} \simeq 0.017$ at the $\Delta(1232)$ resonance~\cite{Mucke:1998mk}, rising to $\xi_\nu^{p\gamma}\simeq 0.07$--$0.11$ at higher $\sqrt{s}$ as determined from \textsc{Pythia\,8} $p\gamma$ simulations (see Appendix~\ref{app:pythia}). For a source of X-ray luminosity $L_X$ with characteristic photon energy $\varepsilon_X$ and emission region of size $R$,
\begin{equation}
N_\gamma^{\rm eff} = \frac{L_X}{4\pi R \, c \, \varepsilon_X}.
\label{eq:Ngamma}
\end{equation}

Lower and upper bounds on $\sigma_{p\gamma}$ follow by the same logic as for $pp$:
\begin{align}
\sigma_{p\gamma} & 
%\gtrsim -\,\frac{1}{\,N_{\gamma,\max}^{\rm eff}}\,
%\ln\!\left(1 - \frac{L_\nu}{\xi_\nu^{p\gamma}\,L_{\rm CR,max}}\right), 
\gtrsim \,\frac{1}{3\, \xi_\nu^{p\gamma} N_{\gamma,\max}^{\rm eff} \,L_{\rm CR,max}} ,
\label{eq:pgamma_lower}
 \\
%\end{equation}
%\begin{equation}
\sigma_{p\gamma} & 
%\lesssim \frac{-\ln\!\left(1 - \dfrac{L_\nu}{\xi_\nu^{p\gamma}\,L_{\rm CR,min}}\right)}{N_{\gamma,\min}^{\rm eff}}.
\lesssim \,\frac{1}{3\, \xi_\nu^{p\gamma} N_{\gamma,\min}^{\rm eff} \,L_{\rm CR,min}} . 
\label{eq:pgamma_upper}
\end{align}
The relevant center-of-mass energy for $p\gamma$ interactions with target photons of energy $\varepsilon_X$ is the isotropic-averaged invariant $\sqrt{s_{p\gamma}} = \sqrt{m_p^2 + 2 E_p \varepsilon_X\,\langle 1-\cos\theta\rangle}$, where for an isotropic target field $\langle 1-\cos\theta\rangle=1$, so that $\sqrt{s_{p\gamma}} = \sqrt{m_p^2 + 2 E_p \varepsilon_X}$; for $E_p = E_\nu / \xi_\nu^{p\gamma}$ (with the energy-dependent $\xi_\nu^{p\gamma}(\sqrt{s})$ from \textsc{Pythia\,8} $p\gamma$ simulations; see Appendix~\ref{app:pythia}) and $\varepsilon_X = 10$~keV this gives $\sqrt{s_{p\gamma}} \sim 1$--$9$~GeV for the IceCube sources. 
For head-on collisions, one would obtain a factor $\sqrt{2}$ larger value; the isotropic approximation is appropriate for coronal and the Central Molecular Zone (CMZ) target fields and constitutes a conservative choice at the $\mathcal{O}(1)$ level. Our prescription is an approximation, and we acknowledge that the X-ray distribution may not be exactly isotropic. 

The photon column density depends on the compactness of the emission region: for AGN coronae, the size is conventionally parametrized as $R = n_{R_g} \, R_g$ with $R_g = G M_{\rm BH}/c^2$ the gravitational radius and $n_{R_g} \sim 3$--$30$~\cite{Fabian:2015MNRAS}. 
For blazar jets, the photon density is inferred from compactness and variability arguments~\cite{Murase:2018iyl, Keivani:2018rnh}. The source-specific parameters used for each source are listed in Table~\ref{tab:sources_transposed}, and the resulting bounds are shown in Fig.~\ref{fig:limits_pgamma}.

Note that the lower bounds on both $\sigma_{pp}$ and $\sigma_{p\gamma}$ are obtained under the assumption 
that the entire neutrino luminosity is produced by inelastic $pp$ or $p\gamma$ scattering, respectively.  It is always possible 
to reduce the $\sigma_{pp,p\gamma}^\mathrm{low}$ by reducing the fraction of the neutrino luminosity produced by either channel.  
But this will have only an ${\cal O}(1)$ effect, which is negligible compared to other uncertainties, unless one of the channels is 
heavily suppressed.  We can thus ignore this effect unless the effect of BSM physics is so dramatic as to cause one channel to be 
almost entirely suppressed.  As increasing numbers of neutrinos sources are found and characterized, the lower bound will become 
more robust.  For example, if the upper bound  $\sigma_i^\mathrm{hi}$ (where $i$ represents a particular channel) from one 
target is too small 
to entirely explain the neutrino luminosity of another target, then one can find a minimum cross section $\sigma^\mathrm{low}$ 
for the other channel 
which does not rely on any assumptions.  Such a global analysis will be an interesting topic of research once more targets have been identified, 
but is premature at this stage.

%=====================================================================
\section{Bounds from the Seyfert population}
\label{sec:seyfert}
%=====================================================================

IceCube has recently reported evidence at the $3.3 \sigma$ level for neutrino emission from a population of X-ray bright AGNs~\cite{IceCube:2024dou}, identified through a binomial test of correlated excess signal from 47 hard X-ray selected AGNs from the \textit{Swift}/BAT catalog~\cite{Baumgartner:2013}. This population signal provides an independent handle on hadronic cross sections through stacking.

We consider the eleven X-ray bright AGNs from the IceCube sample, identified in the recent binomial test analysis~\cite{IceCube:2024dou} as contributing to a $3.3\sigma$ excess (excluding NGC~1068, which is treated separately above). The observed neutrino energies span $E_\nu \simeq 1$--$30$~TeV, corresponding to $\sqrt{s}_{pp} \simeq 145$--$790$~GeV for $pp$ interactions (using the \textsc{Pythia}-derived $\xi_\nu$) and $\sqrt{s}_{p\gamma} \simeq 1.4$--$3.1$~GeV for $p\gamma$ interactions with 10~keV coronal X-ray targets.

For each source, the intrinsic 20--50~keV X-ray luminosity $L_X^{\rm intr}$ is computed directly from the absorption-corrected intrinsic fluxes in the BASS catalog~\cite{Koss:2022ApJS}. Note that for the Seyfert sources, the X-ray luminosity
determines both the photon target density $N_\gamma^{\rm eff}$
and, through coronal acceleration models in which magnetic
reconnection simultaneously heats the plasma and accelerates
nonthermal protons~\cite{Murase:2019vdl, Inoue:2019yfs}, the
cosmic-ray luminosity $L_{\rm CR} = \eta_{\rm CR}\,L_X^{\rm intr}$
with $\eta_{\rm CR} \in [1,10]$.

For $pp$ interactions, the effective proton column density is set by the Thomson optical depth of the corona, $N_p = \tau_T / \sigma_T$, with $\tau_T \in [1, 3]$~\cite{Fabian:2015MNRAS, Murase:2019vdl}. This gives $N_p^{\rm eff} \in [1.5, 4.5] \times 10^{24}$~cm$^{-2}$, common to all sources. For $p\gamma$ interactions, the photon column density $N_\gamma^{\rm eff}$ depends on the individual source properties ($L_X^{\rm intr}$, $M_{\rm BH}$) and the assumed corona size $n_{R_g} \in [3, 30]$. The stacked effective photon column is computed as a luminosity-weighted average, $N_\gamma^{\rm eff} = \sum_i L_{X,i}^{\rm intr} N_{\gamma,i} / \sum_i L_{X,i}^{\rm intr}$.

The stacked neutrino luminosity, estimated from the population-level best-fit flux (accounting for the fact that individual 90\% C.L.\ upper limits overestimate the actual population signal when summed), is $L_\nu^{\rm stack} \simeq 2.3 \times 10^{43}$~erg~s$^{-1}$ and the total intrinsic X-ray luminosity is $L_X^{\rm intr, stack} = 8.5 \times 10^{44}$~erg~s$^{-1}$. 
Applying Eqs.~(\ref{eq:sigma_pp}) 
and~(\ref{eq:pgamma_lower})--(\ref{eq:pgamma_upper}) with these stacked quantities gives bands of allowed cross-section values. For $pp$, the band spans $\sigma_{pp} \in [2.3,\, 70]$~mb; the SM expectation of $\sim 40$--$70$~mb sits comfortably within this range, and the upper bound is significantly tighter than that from NGC~1068 alone ($\sim 1333$~mb) owing to the energetic constraint from the larger stacked neutrino luminosity. For $p\gamma$, the band spans $\sigma_{p\gamma} \in [0.08,\, 9]$~mb. The SM value of $\sim 0.15$--$0.3$~mb lies within the allowed band, reflecting the relatively low photon column densities in Seyfert coronae that make $p\gamma$ interactions subdominant to $pp$. The full set of per-source parameters is given in Table~\ref{tab:seyfert_inputs}, and the stacked bands are shown in Figs.~\ref{fig:limits_pp} and~\ref{fig:limits_pgamma}.

%=====================================================================
\section{Projected bounds}
\label{sec:projected}
%=====================================================================

\paragraph*{KM3-230213A.---}
The ultra-high-energy neutrino event detected by KM3NeT at $E_\nu \simeq 220$~PeV~\cite{KM3NeT:2025npi} would, if associated with an AGN, probe $pp$ cross sections at $\sqrt{s} \simeq (4.8$--$9.8) \times 10^4$~GeV and $p\gamma$ cross sections at $\sqrt{s} \simeq 155$--$316$~GeV---energies beyond the reach of HERA. In the absence of a confirmed source association we present the KM3-230213A projection under a ``standard-candle'' idealization: we fix the target column and the CR luminosity to their best-estimate values and propagate only the neutrino-luminosity uncertainty. This isolates the intrinsic sensitivity of the measurement and produces bands centred on the SM prediction at the relevant $\sqrt{s}$. For $pp$ we assume a column density $N_p^{\rm eff} \sim 10^{24.8}$--$10^{25.7}$~cm$^{-2}$ and a CR-to-neutrino luminosity ratio $L_{\rm CR}/L_\nu \sim 20$, which yields an allowed band of $\sigma_{pp} \sim 28$--$286$~mb, symmetric around the SM value of $\sim 100$~mb at $\sqrt{s} \sim 50$--$100$~TeV. For $p\gamma$, the standard-candle idealization gives a narrow band $\sigma_{p\gamma} \simeq 0.06$--$0.38$~mb centred on the SM expectation of $\sim 0.15$~mb at $\sqrt{s} \sim 155$--$316$~GeV. The resulting projected bounds are shown in Figs.~\ref{fig:limits_pp} and~\ref{fig:limits_pgamma}.

\paragraph*{Ultra-high-energy $p\gamma$ prospects: GZK neutrinos from X-ray dense sources.---}
Perhaps the most striking projected sensitivity involves EeV-scale neutrinos from X-ray bright AGN. At these energies, CR protons accelerated to GZK-scale energies ($E_p \gtrsim 10^{18}$~eV) interact with the dense $\sim 10$~keV X-ray photon field of an AGN corona, producing neutrinos via $p\gamma \to \pi X$. The resulting center-of-mass energy is
\begin{equation}
\sqrt{s_{p\gamma}} = \sqrt{m_p^2 + 2 E_p \varepsilon_X} \sim 140\text{--}1400~\text{GeV},
\end{equation}
for $E_p \sim 10^{18}$--$10^{20}$~eV and $\varepsilon_X = 10$~keV. This energy range lies far beyond the reach of HERA ($\sqrt{s} \simeq 209$~GeV) and fixed-target photoproduction experiments, probing the $p\gamma$ cross section in an entirely unexplored regime. We show this projected sensitivity in Fig.~\ref{fig:limits_pgamma} as the ``UHE $p\gamma$ (opt.)'' band, which represents the most optimistic configuration: perfect knowledge of the neutrino luminosity, perfect knowledge of the target photon column density, and restriction to the highest attainable $\sqrt{s}$ portion of the accessible energy range ($\sqrt{s} \sim 0.5$--$1.5$~TeV). The resulting band, $\sigma_{p\gamma} \simeq 0.12$--$0.35$~mb, is a factor of $\sim 2$ wide around the SM prediction at these energies, illustrating the ultimate reach of the neutrino-telescope approach once astrophysical uncertainties are under control. Such a detection could come from next-generation neutrino observatories including IceCube-Gen2~\cite{IceCube-Gen2:2020qha}, GRAND~\cite{GRAND:2018iaj}, or the radio component of the Trinity telescope~\cite{Otte:2018uxj}, which are expected to achieve sensitivity to neutrino fluxes at $E_\nu > 10^{17}$~eV. The combination of GZK-scale CR acceleration and dense coronal X-ray fields thus offers a unique window into photoproduction physics at center-of-mass energies complementary to future collider proposals.

\paragraph*{Cosmogenic neutrinos.---}
Future observations of cosmogenic neutrinos at ultra-high energies with radio telescopes would extend the sensitivity to the largest center-of-mass energies. Cosmogenic neutrinos arise from CR interactions with the cosmic microwave background (CMB) and extragalactic background light (EBL), whose number densities are much better constrained than those within individual astrophysical sources.

The CMB photon number density is fixed by the precisely measured temperature,
\begin{equation}
n_\gamma = \frac{2\zeta(3)}{\pi^2} T_{\rm CMB}^3,
\end{equation}
with a fractional uncertainty of only $\delta n_\gamma / n_\gamma \simeq 6.3 \times 10^{-4}$ from the FIRAS measurement $T_{\rm CMB} = 2.72548 \pm 0.00057$~K~\cite{Fixsen_2009}. The EBL photon field is less precisely known, with current constraints at the level $\delta n_{\rm EBL}/n_{\rm EBL} \sim \mathcal{O}(10\%)$ from \textit{Fermi}-LAT and imaging atmospheric Cherenkov telescope data~\cite{MAGIC:2019ozu}. As a result, uncertainties in cosmogenic neutrino flux predictions are dominated by the ultra-high-energy CR spectrum, composition, and source evolution, rather than by the target photon densities. The effective cosmological photon column $N_\gamma^{\rm eff} \sim 10^{30}$--$10^{31}$~cm$^{-2}$~\cite{Fixsen_2009, 2011MNRAS.410.2556D} and the well-measured UHECR flux make this a uniquely clean environment for constraining $p\gamma$ cross sections at $\sqrt{s} \sim 0.1$--$10$~GeV.

\paragraph*{CR--CR collisions.---}
A brief analysis of CR--CR collisions shows that, under known physics, this channel is negligible compared to CR--gas interactions: even aggressive power-law growths $\sigma \propto (\sqrt{s})^p$ with $p\sim 1.5$--$3$ anchored at the SM value at $\sqrt{s}_{\rm GZK}$\footnote{$\sqrt{s}_{\rm GZK}$ denotes the center-of-mass energy of a GZK-scale proton ($E \sim 10^{20}$~eV) colliding with a proton at rest.} remain below the Froissart--Martin unitarity bound out to the kinematic limit. This analysis, together with Fig.~\ref{fig:limits_CRCR}, is presented in Appendix~\ref{app:crcr} and validates the CR--gas treatment adopted in the main text.

\begin{figure*}[t!]
\centering
\includegraphics[width=0.8\textwidth]{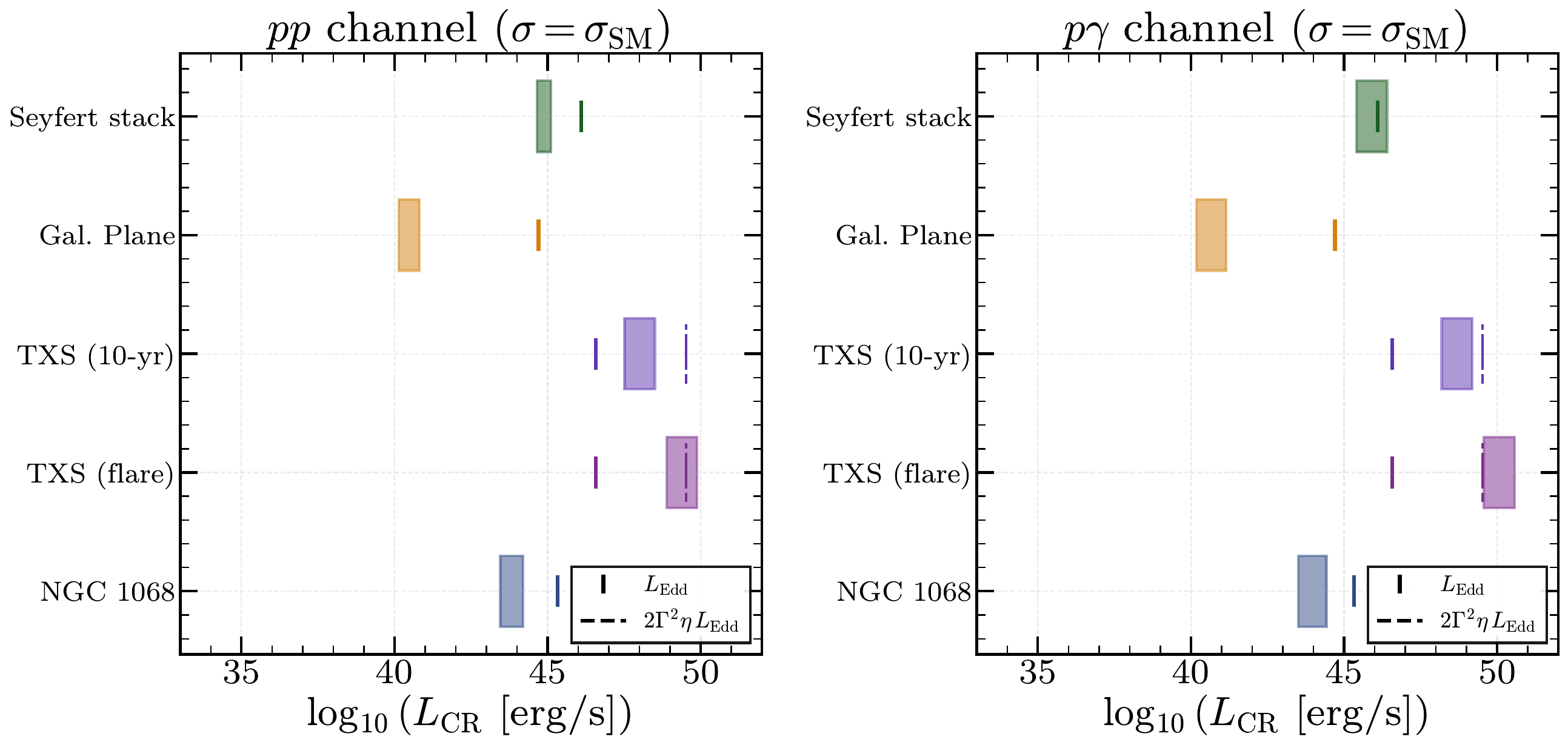}
\caption{\textbf{\textit{CR luminosity required to reproduce the measured
neutrino luminosity when the cross section is fixed to the
SM prediction, for both the $pp$ (left) and $p\gamma$ (right) channels.}}
Shaded bands reflect the range of
effective target column densities for each source
(Table~\ref{tab:sources_transposed}). The Seyfert-stack entry uses
the luminosity-weighted average black-hole mass of the 11 contributing
Seyfert galaxies. TXS~0506+056 is shown separately for the 2014--15 flare and the 10-year time-integrated luminosity. Solid vertical tick marks indicate
the Eddington luminosity $L_{\rm Edd}$ of the central black hole; dashed tick marks show the beaming-corrected limit $2\Gamma^2\eta\,L_{\rm Edd}$ for TXS entries ($\Gamma=15$, $\eta=2$). Sources whose
CR band exceeds $L_{\rm Edd}$ require super-Eddington CR power, though beaming corrections can relax this tension.}
\label{fig:CR_luminosity}
\end{figure*}

\paragraph*{Bounding the CR Luminosity.---}
An alternative and complementary perspective is obtained by
inverting the problem: rather than bounding $\sigma_{pp}$, one
can fix the cross section to its SM value and instead
derive the CR luminosity required to explain the
observed neutrino flux. Since $L_\nu = 3\,\xi_\nu\,f_{pp}\,L_{\rm CR}$,
the measured $L_\nu$ directly translates into a required
$L_{\rm CR}$ once $\sigma_{pp}^{\rm SM}$ and $N_p^{\rm eff}$
are specified. The resulting constraints are shown in
Fig.~\ref{fig:CR_luminosity}. For each source, the shaded band
spans the range of effective target column densities listed in
Table~\ref{tab:sources_transposed}; TXS~0506+056 is shown separately
for the 2014--15 flare and the 10-year average. Solid vertical tick marks indicate
the Eddington luminosity
$L_{\rm Edd} = 4\pi\,G\,M_{\rm BH}\,m_p\,c/\sigma_T$;
for TXS, dashed tick marks show the beaming-corrected limit
$2\Gamma^2\eta\,L_{\rm Edd}$ ($\Gamma=15$, $\eta=2$).
Sources whose required CR luminosity exceeds $L_{\rm Edd}$
are in apparent tension with the Eddington limit, though
relativistic beaming can relax this requirement, providing a self-consistency check on the
astrophysical parameters adopted.

%=====================================================================
\section{Laboratory and Atmospheric data, and the Froissart--Martin bound}
\label{sec:laboratory}
%=====================================================================

\subsection{Origin of the $pp$ reference data}

The grey data points shown in Fig.~\ref{fig:limits_pp} (and, in the appendix, Fig.~\ref{fig:limits_CRCR}) represent the world dataset of $pp$ (and, at lower energies, $p\bar{p}$) inelastic cross-section measurements compiled by the Particle Data Group (PDG)~\cite{ParticleDataGroup:2024cfk}. These span $\sqrt{s} \simeq 2$--$95\,000$~GeV and originate from several experimental programs, each with distinct systematics:

\emph{Fixed-target and ISR era} ($\sqrt{s} \lesssim 63$~GeV). Early measurements at the CERN Proton Synchrotron and the Intersecting Storage Rings achieved $1$--$2\%$ precision using missing-mass and forward-scattering techniques~\cite{CERN-Pisa-Rome-StonyBrook:1976mtu}. The dominant systematic is the correction for single-diffraction events, which depends on the definition of ``inelastic''.

\emph{S$p\bar{p}$S and Tevatron} ($\sqrt{s} = 546$--$1800$~GeV). The UA5, CDF, and E811 experiments measured $p\bar{p}$ total and inelastic cross sections with $\sim 2$--$3\%$ uncertainty~\cite{E811:1998got}. The $p\bar{p}$ inelastic cross section is related to $pp$ via isospin arguments at these energies: Regge theory predicts convergence of $pp$ and $p\bar{p}$ cross sections at $\sqrt{s} \gtrsim 100$~GeV; below this energy, the difference is $\lesssim 5$~mb.

\emph{LHC} ($\sqrt{s} = 7$--$13$~TeV). ATLAS~\cite{ATLAS:2016ygv}, CMS~\cite{CMS:2012zex}, TOTEM~\cite{TOTEM:2017asr}, and ALICE~\cite{ALICE:2012xs} measured $\sigma_{pp}^{\rm inel}$ at $\sqrt{s} = 7$, $8$, and $13$~TeV with combined uncertainties of $1.5$--$2.5\%$. The dominant systematics are the extrapolation to low diffractive masses and the luminosity calibration.

\emph{CR air showers} ($\sqrt{s} \sim 10^4$--$10^5$~GeV). The Pierre Auger Observatory~\cite{PierreAuger:2012egl} and the Telescope Array~\cite{TelescopeArray:2015oxb} inferred $\sigma_{pp}^{\rm inel}$ at $\sqrt{s} \sim 57$ and $95$~TeV from the distribution of the atmospheric depth of shower maximum, $X_{\rm max}$. The conversion from the measured proton--air cross section to $pp$ relies on hadronic interaction models (e.g., EPOS-LHC, QGSJet-II, Sibyll), introducing model-dependent uncertainties of $\sim 10$--$20\%$~\cite{Ulrich:2010rg}.
Crucially, the Auger measurement is independent of the absolute CR flux: the proton--air cross section is extracted from the shape of the $X_{\rm max}$ distribution---specifically the exponential slope $\Lambda_\eta$ of its deep tail, which is determined by the interaction length $\lambda_{\rm int}=\langle m_{\rm air}\rangle/\sigma_{p\text{-air}}$ and is insensitive to how many showers are observed~\cite{PierreAuger:2012egl}. This stands in contrast to the neutrino-telescope approach presented here, which is inherently flux-based: the observed neutrino luminosity encodes the product $\xi_\nu\,f_{pp}\,L_{\rm CR}$, so isolating $\sigma_{pp}$ requires independent knowledge of the CR luminosity and target column density. The two methods are therefore complementary: Auger exploits a shape-based observable that cancels the flux normalization, while neutrino telescopes exploit the absolute luminosity but require the astrophysical parameters to be bounded by independent electromagnetic data and theoretical arguments.

\subsection{Origin of the $p\gamma$ reference data}

The reference data in Fig.~\ref{fig:limits_pgamma} are from the PDG compilation of total $p\gamma$ cross-section measurements~\cite{ParticleDataGroup:2024cfk}. At $\sqrt{s} \lesssim 20$~GeV, fixed-target photoproduction experiments at SLAC, DESY, and Fermilab dominate the dataset, with typical uncertainties of $2$--$5\%$ arising from photon flux normalization and detector acceptance corrections. At $\sqrt{s} = 20$--$209$~GeV, the H1 and ZEUS experiments at the HERA $ep$ collider measured $\sigma_{p\gamma}^{\rm tot}$ by tagging quasi-real photons ($Q^2 \to 0$) in deep-inelastic scattering events~\cite{H1:2000muc, ZEUS:2001ixs}. These measurements reach $\sqrt{s} \simeq 209$~GeV with $\sim 5$--$10\%$ precision; the primary systematics are the photon flux factor and the extrapolation to the photoproduction limit $Q^2 \to 0$. No laboratory data exist beyond $\sqrt{s} \simeq 209$~GeV, making the neutrino-telescope constraints on $\sigma_{p\gamma}$ at higher energies the first probes of this regime.

\subsection{Froissart--Martin unitarity bound}

A theoretical upper limit on hadronic cross sections follows from unitarity and analyticity of the $S$-matrix. The scattering amplitude can be expanded in partial waves,
\begin{equation}
A(s,t) = 16\pi \sum_{\ell=0}^{\infty} (2\ell+1) \, a_\ell(s) \, P_\ell(\cos\theta),
\end{equation}
where unitarity requires $|a_\ell(s)| \le 1$. The optical theorem relates the total cross section to the forward amplitude,
\begin{equation}
\sigma_{\rm tot}(s) = \frac{4\pi}{k^2} \sum_{\ell=0}^{\infty} (2\ell+1) \, \mathrm{Im}\, a_\ell(s),
\end{equation}
where $k$ is the center-of-mass momentum. Causality and the mass gap set by the pion restrict the number of significant partial waves, so that if all contributing waves saturate the unitarity condition, one obtains the Froissart--Martin bound~\cite{Froissart:1961ux, Martin:1965jj, Martin:2009pt},
\begin{equation}
\sigma_{\rm FM}(s) = \frac{\pi}{m_\pi^2} \ln^2\!\frac{s}{s_0},
\end{equation}
where $s_0$ is a hadronic scale set by convention, and a factor of four smaller limit for the inelastic component~\cite{Martin:2009pt}. These are shown in Fig.~\ref{fig:limits_pp} as dotted and dashed lines, respectively. Notably, the bounds inferred from IceCube data are more stringent than the unitarity bound at several center-of-mass energies.

\subsection{Complementarity with neutrino-telescope bounds}

For the $pp$ channel, the neutrino-telescope bounds cover $\sqrt{s} \sim 100$--$10^5$~GeV, overlapping with the LHC regime at the lower end and extending towards the highest-energy CR measurements at the upper end (via KM3-230213A). In the overlap region ($\sqrt{s} \sim 100$--$10^4$~GeV), the neutrino bounds are wider than laboratory measurements but provide an independent cross-check from a completely different experimental methodology. At several center-of-mass energies unscanned by colliders, neutrino telescopes provide the only available constraints.

For the $p\gamma$ channel, the projected KM3-230213A and UHE constraints probe $\sqrt{s} \sim 100$--$300$~GeV, extending well beyond the HERA limit. A future detection of an EeV-scale neutrino from an X-ray bright AGN would access an entirely unexplored energy regime for photoproduction cross sections beyond $\sqrt{s} \gtrsim 10^{3}$ GeV

The importance of the neutrino--collider complementarity extends beyond $pp$ and $p\gamma$ interactions. There is a large literature on probing new physics through deviations from the SM deep-inelastic neutrino--nucleon cross section at center-of-mass energies comparable to collider probes~\cite{Alvarez-Muniz:2001efi, Alvarez-Muniz:2002snq, Albuquerque:2003mi, Anchordoqui:2006wc, Herrera:2025pdn, Bai:2025pef,Palmisano:2026sid}. What has perhaps received less attention is the fact that neutrino telescopes also bracket the $pp$ and $p\gamma$ cross sections themselves, at energies beyond those of terrestrial accelerators.

%--- Table I: Sources and assumptions ---
\begin{table*}[t!]
\centering
\caption{\textbf{\textit{Sources and assumptions used to derive bounds on the inelastic $pp$ and $p\gamma$ cross sections from neutrino observations.}} When two entries are given, quantities are quoted as $pp$ / $p\gamma$. For $pp$ interactions, the target proton is assumed at rest; for $p\gamma$ interactions, the characteristic target photon energy is indicated.}
\label{tab:sources_transposed}
\small
\setlength{\tabcolsep}{4.5pt}
\renewcommand{\arraystretch}{1.18}
\resizebox{\textwidth}{!}{%
\begin{tabular}{lcccccc}
\hline\hline
 & TXS~0506+056 & NGC~1068 & Gal.~plane & KM3-230213A & Transient & Cosmogenic \\
\hline
Detector
& IceCube & IceCube & IceCube & KM3NeT & Radio & Radio \\
$E_\nu$ [eV]
& $4\times10^{13}$--$4\times10^{14}$
& $1.5\times10^{12}$--$3\times10^{13}$
& $10^{12}$--$10^{14}$
& $(1.2$--$5)\times10^{17}$
& $\gtrsim10^{17}$
& $\gtrsim10^{17}$ \\
Target
& $p/\gamma(10~{\rm keV})$
& $p/\gamma(10~{\rm keV})$
& $p/\gamma(10~{\rm keV})$
& $p/\gamma(10~{\rm keV})$
& $\gamma(10~{\rm keV})$
& $\gamma(10^{-3}$--$1~{\rm eV})$ \\
$\sqrt{s}_{pp}$ [GeV]${}^{*}$
& $0.9\times10^{3}$--$2.8\times10^{3}$
& $1.8\times10^{2}$--$7.9\times10^{2}$
& $1.5\times10^{2}$--$1.4\times10^{3}$
& $(4.8$--$9.8)\times10^{4}$
& ---
& --- \\
$\sqrt{s}_{p\gamma}$ [GeV]
& $3.5$--$10$
& $1.4$--$3.1$
& $1.4$--$5.4$
& $151$--$311$
& $141$--$1414$
& $10^{-1}$--$10^{1}$ \\
$N^{\rm eff}_p$ [cm$^{-2}$]
& $10^{23}$--$10^{24}{}^{a}$
& $1.5\times10^{24}$--$10^{25}{}^{b}$
& $6\times10^{23}$--$3\times10^{24}{}^{c}$
& $7\times10^{24}$--$5\times10^{25}$
& ---
& --- \\
$N^{\rm eff}_\gamma$ [cm$^{-2}$]
& $10^{25}$--$10^{26}{}^{a'}$
& $3\times10^{26}$--$3\times10^{27}{}^{b'}$
& $10^{26}$--$10^{27}{}^{c'}$
& $5\times10^{26}$--$5\times10^{27}$
& ---
& $10^{30}$--$10^{31}{}^{d}$ \\
$L_{\rm CR}$ [erg\,s$^{-1}$]
& $(2.6\times10^{48}$--$3.4\times10^{49}){}^{e}$
& $5\times10^{43}$--$10^{45}{}^{f}$
& $5\times10^{40}$--$2\times10^{41}{}^{g}$
& $(3$--$20)\,L_\nu{}^{h}$
& ---
& UHECR \\
$L_\nu^{\rm meas}$ [erg\,s$^{-1}$]
& $(1.2^{+0.6}_{-0.4})\times10^{47}{}^{i}$
& $(2.9\pm0.9)\times10^{42}{}^{j}$
& $5\times10^{38}{}^{k}$
& ---$^{\ell}$
& ---
& --- \\
\hline\hline
\end{tabular}%
}
\begin{flushleft}
\footnotesize
$^{a}$\,TXS $N_p^{\rm eff}$ from variability and transparency arguments~\cite{Murase:2018iyl,Celotti:2007rb}.\\
$^{a'}$\,Effective X-ray photon column in the TXS~0506+056 jet from compactness and variability arguments, assuming a 10~keV target~\cite{Murase:2018iyl,Keivani:2018rnh}.\\
$^{b}$\,NGC~1068: lower end from coronal Thomson depth $\tau_T \sim 1$; upper end from the Compton-thick column density $N_H \sim 10^{25}$~cm$^{-2}$ measured in X-ray spectroscopy~\cite{Marinucci:2015fqo,Bauer:2014rla}.\\
$^{b'}$\,Effective X-ray photon column density in NGC~1068, consistent with corona-scale photon densities~\cite{Murase:2022dog,Inoue:2019yfs}.\\
$^{c}$\,Galactic-plane effective proton column density from B/C grammage measurements at TeV energies~\cite{AMS:2016brs,CALET:2022vro}.\\
$^{c'}$\,X-ray photon column density from diffuse Galactic X-ray emission~\cite{Ponti:2015tva,Mori:2015vba}.\\
$^{d}$\,Effective cosmological photon column (CMB/EBL)~\cite{Fixsen_2009,2011MNRAS.410.2556D}.\\
$^{e}$\,TXS CR power from BZ+MAD beaming: $L_{\rm CR,max}^{\rm iso} = 2\Gamma_{\rm jet}^2\,\eta_{\rm jet}\,L_{\rm Edd}$ with $\Gamma_{\rm jet}=15$, $\eta_{\rm jet}=2$ (see text).\\
$^{f}$\,NGC~1068 CR luminosity bracketed by the intrinsic X-ray luminosity from below and $\sim L_{\rm bol}/2$ from above, consistent with AGN corona neutrino models~\cite{Murase:2022dog,Inoue:2019yfs}.\\
$^{g}$\,Galactic CR power from SN energetics: $\sim 2$--$3$ SN/century $\times 10^{51}$~erg $\times 10\%$ efficiency~\cite{Strong:2010pr}.\\
$^{h}$\,Upper limit from kinematics; CR luminosity is uncertain without source association.\\
$^{i}$\,IceCube time-dependent TXS flare (2014--2015),
isotropic-equivalent all-flavor $L_\nu$~\cite{IceCube:2018cha}.
The time-averaged luminosity is substantially lower; we adopt
the flare value as it yields the most constraining bounds.\\
$^{k}$\,All-flavor, model-independent neutrino luminosity from the IceCube Galactic-plane detection~\cite{IceCube:2023ame}.\\
$^{\ell}$\,KM3NeT event-level detection; converting to $L_\nu$ requires source association~\cite{KM3NeT:2025npi}.\\
$^{*}$\,Center-of-mass energies computed using the energy-dependent $\xi_\nu(\sqrt{s})$ from \textsc{Pythia\,8} simulations (see Appendix and Figs.~\ref{fig:pythia_dist},~\ref{fig:pythia_props}).
\end{flushleft}
\end{table*}

%--- Table II: Seyfert population ---
\begin{table*}[t!]
\centering
\caption{\textbf{\textit{Input parameters for the cross-section constraints from the eleven X-ray bright AGN identified by IceCube~\cite{IceCube:2024dou} (excluding NGC~1068) and their stacked combination.}} Intrinsic 20--50~keV X-ray fluxes are taken from the BASS catalog~\cite{Koss:2022ApJS}, black hole masses from~\cite{Koss:2022ApJS}. Neutrino luminosities are 90\% C.L. upper limits derived from the IceCube 90\% flux upper limits assuming a $\gamma=3$ spectrum.}
\label{tab:seyfert_inputs}
\smallskip
\resizebox{\textwidth}{!}{%
\begin{tabular}{l c c c c c c c c c c c c}
\hline\hline
 & NGC 7469 & NGC 4151 & CGCG 420-015 & Cygnus A & LEDA 166445 & NGC 4992 & NGC 1194 & Mrk 1498 & MCG +4-48-2 & NGC 3079 & Mrk 417 & \textbf{Stacked} \\
\hline
Detector & \multicolumn{12}{c}{IceCube~\cite{IceCube:2024dou}} \\[2pt]
$E_\nu$ & \multicolumn{12}{c}{$[1,\,30]$~TeV} \\[2pt]
Target ($pp$) & \multicolumn{12}{c}{Coronal protons ($\tau_T \in [1,3]$)~\cite{Fabian:2015MNRAS,Murase:2019vdl}} \\[2pt]
Target ($p\gamma$) & \multicolumn{12}{c}{Coronal X-rays ($\varepsilon_X = 10$~keV, $n_{R_g} \in [3,30]$)} \\[2pt]
$\sqrt{s}_{pp}$ [GeV]${}^{*}$ & \multicolumn{12}{c}{$[145,\,790]$} \\[2pt]
$\sqrt{s}_{p\gamma}$ [GeV] & \multicolumn{12}{c}{$[1.0,\,2.6]$} \\[2pt]
$N_p^{\rm eff}$ [cm$^{-2}$] & \multicolumn{12}{c}{$[1.5,\, 4.5] \times 10^{24}$\,$^a$} \\[4pt]
\hline
$N_\gamma^{\rm eff}$ [cm$^{-2}$]\,$^b$
 & $[5.0 \times 10^{25}$, & $[6.4 \times 10^{24}$, & $[1.3 \times 10^{25}$, & $[5.7 \times 10^{24}$, & $[1.8 \times 10^{25}$, & $[1.2 \times 10^{25}$, & $[8.9 \times 10^{24}$, & $[4.4 \times 10^{25}$, & $[6.2 \times 10^{25}$, & $[2.7 \times 10^{25}$, & $[9.0 \times 10^{24}$, & $[2.4 \times 10^{25}$, \\
 & $5.0 \times 10^{26}]$ & $6.4 \times 10^{25}]$ & $1.3 \times 10^{26}]$ & $5.7 \times 10^{25}]$ & $1.8 \times 10^{26}]$ & $1.2 \times 10^{26}]$ & $8.9 \times 10^{25}]$ & $4.4 \times 10^{26}]$ & $6.2 \times 10^{26}]$ & $2.7 \times 10^{26}]$ & $9.0 \times 10^{25}]$ & $2.4 \times 10^{26}]$\,$^c$ \\[4pt]
$L_{\rm CR}$ [erg\,s$^{-1}$]\,$^d$
 & $[1.6 \times 10^{43}$, & $[7.8 \times 10^{42}$, & $[3.5 \times 10^{43}$, & $[3.8 \times 10^{44}$, & $[4.8 \times 10^{43}$, & $[3.3 \times 10^{43}$, & $[1.6 \times 10^{43}$, & $[1.2 \times 10^{44}$, & $[1.7 \times 10^{44}$, & $[1.7 \times 10^{42}$, & $[2.4 \times 10^{43}$, & $[8.5 \times 10^{44}$, \\
 & $1.6 \times 10^{44}]$ & $7.8 \times 10^{43}]$ & $3.5 \times 10^{44}]$ & $3.8 \times 10^{45}]$ & $4.8 \times 10^{44}]$ & $3.3 \times 10^{44}]$ & $1.6 \times 10^{44}]$ & $1.2 \times 10^{45}]$ & $1.7 \times 10^{45}]$ & $1.7 \times 10^{43}]$ & $2.4 \times 10^{44}]$ & $8.5 \times 10^{45}]$ \\[4pt]
$L_\nu^{\rm 90\% UL}$ [erg\,s$^{-1}$]
 & $7.4 \times 10^{43}$ & $3.6 \times 10^{42}$ & $2.0 \times 10^{44}$ & $5.5 \times 10^{44}$ & $1.9 \times 10^{44}$ & $1.1 \times 10^{44}$ & $3.4 \times 10^{43}$ & $3.6 \times 10^{44}$ & $2.3 \times 10^{44}$ & $2.8 \times 10^{42}$ & $8.3 \times 10^{43}$ & $2.3 \times 10^{43}$\,$^e$ \\[2pt]
\hline\hline
\end{tabular}%
}
\begin{flushleft}
\footnotesize
$^a$\,$N_p = \tau_T / \sigma_T$ with $\tau_T \in [1,3]$. Common to all sources.\\
$^b$\,$N_\gamma = L_X^{\rm intr}/(4\pi R\,c\,\varepsilon_X)$ with $R = n_{R_g}\,R_g$, $R_g = GM_{\rm BH}/c^2$, and $n_{R_g} \in [3,30]$ (extended to compact). Intrinsic luminosity $L_X^{\rm intr}$ computed from the BASS 20--50~keV intrinsic X-ray flux.\\
$^c$\,Luminosity-weighted effective: $N_\gamma^{\rm eff} = \sum_i L_{X,i}^{\rm intr}\,N_{\gamma,i}\,/\,\sum_i L_{X,i}^{\rm intr}$.\\
$^d$\,$L_{\rm CR} = \eta_{\rm CR}\,L_X^{\rm intr}$ with $\eta_{\rm CR} \in [1,10]$. Stacked column uses $L_X^{\rm intr,stack} = 8.45 \times 10^{44}$~erg\,s$^{-1}$.\\
$^e$\,Stacked neutrino luminosity is estimated from the population-level best-fit flux~\cite{IceCube:2024dou}, not from the sum of individual upper limits (which would overestimate the actual population signal). Individual values are all-flavor luminosities from 90\% C.L. flux upper limits ($\gamma=3$, 1--30~TeV).\\
$^{*}$\,Center-of-mass energies computed using the \textsc{Pythia\,8}-derived $\xi_\nu(\sqrt{s})$ (see Appendix).\\
\end{flushleft}
\end{table*}

%=====================================================================
\section{Conclusions}
\label{sec:conclusions}
%=====================================================================

We have presented a recipe for extracting two-sided bounds on the inelastic $pp$ and $p\gamma$ cross sections from neutrino point-source data, by independently constraining every astrophysical input---CR luminosities and target column densities---through electromagnetic observations or theoretical arguments, leaving the cross section as the only remaining free parameter.

Applying this framework to the IceCube associations with TXS~0506+056, NGC~1068, and the Galactic Plane, to the stacked population of eleven X-ray bright Seyfert galaxies, to the ultra-high-energy KM3NeT event KM3-230213A, and to projected observations of ultra-high-energy neutrinos, we obtain allowed bands for both cross sections that span center-of-mass energies from $\sqrt{s} \sim 100$~GeV to $\sim$~EeV.  Several of these bounds are more stringent than the Froissart--Martin unitarity limit, and the $p\gamma$ constraints extend beyond the kinematic reach of HERA.  While the astrophysical uncertainties are significant---arising primarily from the proton and photon column densities and the intrinsic CR luminosity---the SM cross section is contained within the allowed region for every source, providing a non-trivial consistency check on both the particle physics and the astrophysical modeling.

For TXS~0506+056, adopting a BZ+MAD beaming-corrected CR luminosity ceiling yields finite two-sided bounds for both the 2014--15 flare and the 10-year time-integrated luminosity; the latter provides the tighter constraint owing to its factor $\sim 25$ lower $L_\nu$.  The stacked Seyfert population provides a complementary constraint at $\sqrt{s} \sim 150$--$800$~GeV with the smallest astrophysical uncertainties among the sources considered, owing to the averaging over eleven independent lines of sight.  Inverting the analysis; fixing $\sigma$ to the SM prediction and solving for the required CR luminosity; replace semicolons with commas? Otherwise, the sentence doesn't make sense. highlights that the TXS-flare requires super-Eddington CR power (even after accounting for a beaming correction), while the other sources are well below the Eddington limit.

The detection of additional ultra-high-energy neutrinos by the growing landscape of next-generation observatories will progressively extend these constraints to $\sqrt{s} \sim$~EeV and beyond ~\cite{Arguelles:2024ncf}. Water- and ice-based Cherenkov telescopes---IceCube~\cite{IceCube:2016zyt}, KM3NeT~\cite{KM3Net:2016zxf}, P-ONE~\cite{P-ONE:2020ljt}, Baikal-GVD~\cite{Avrorin:2015wba}, and TRIDENT~\cite{TRIDENT:2022hql}---will continue to identify and characterize individual sources in the TeV--PeV regime. At higher energies, complementary techniques become effective: Earth-skimming tau-neutrino detection at TAMBO~\cite{TAMBO:2025jio}, optical Cherenkov observations from mountain-top arrays~\cite{Otte:2025dld}, and radio detection of in-ice and in-air showers by GRAND~\cite{GRAND:2018iaj, GRAND:2025rps}, RNO-G~\cite{RNO-G:2020rmc}, PUEO~\cite{PUEO:2020bnn}, and BEACON~\cite{BEACON:2021fpe} will deliver the EeV-scale sensitivity required to test QCD and the foundational principles of unitarity and analyticity at energies inaccessible to any terrestrial collider.

The central message of this work is that high-energy neutrino sources operate as hadronic colliders, scanning continuously over large ranges of center-of-mass energies, that in several cases extend well beyond the reach of any terrestrial facility. The width of the allowed cross-section bands is set not primarily by the neutrino data themselves but by astrophysical uncertainties, namely, the cosmic-ray luminosity and target column density. As these inputs are sharpened by deeper multi-messenger and multi-wavelength observations, we foresee neutrino telescopes maturing into precision probes of hadronic physics at energies no Earth-based experiment can access otherwise.

\section*{Acknowledgments}

We are grateful to Aparajitha Karthikeyan, Deepak Sathyan, and Louis Strigari for useful discussions. The work of GH is supported by the Neutrino Theory Network Fellowship with contract number 726844.
CAA are supported by the Faculty of Arts and Sciences of Harvard University, Canadian Institute for Advanced Research (CIFAR), the National Science Foundation (NSF), the John Templeton Foundation, the Research Corporation for Science Advancement, and the David \& Lucile Packard Foundation. The work of PSBD is partly supported by the U.S. Department of Energy under grant No.~DE-SC0017987 and by a Humboldt Fellowship from the Alexander von Humboldt Foundation. The work of BD and MR is supported by the U.S. DOE Grant DE-SC0010813.
The work of JK is supported by U.S.~DOE grant DE-SC0010504. IMS is supported by the U.S. Department of Energy under the award number DE-SC0020262. 
N.K. is supported by the David and Lucile Packard Foundation and NSF IAIFI.
This work is supported by the National Science Foundation under Cooperative Agreement PHY-2019786 (The NSF AI Institute for Artificial Intelligence and Fundamental Interactions, http://iaifi.org/). For facilitating portions of this research, the authors wish to acknowledge the Center for Theoretical Underground Physics and Related Areas (CETUP*), The Institute for Underground Science at Sanford Underground Research Facility (SURF), and the South Dakota Science and Technology Authority for hospitality and financial support, as well as for providing a stimulating environment.

\bibliography{references}
%\clearpage

%%%%%% SUPPLEMENTAL MATERIAL STARTS HERE
%%%%%% SUPPLEMENTAL MATERIAL STARTS HERE
%%%%%% SUPPLEMENTAL MATERIAL STARTS HERE
%%%%%% SUPPLEMENTAL MATERIAL STARTS HERE

%\newpage

\onecolumngrid
\appendix

% \ifx \standalonesupplemental\undefined
% \setcounter{page}{1}
% \setcounter{figure}{0}
% \setcounter{table}{0}
% \setcounter{equation}{0}
% \fi

% \renewcommand{\thepage}{Supplemental Methods and Tables -- S\arabic{page}}
% \renewcommand{\figurename}{SUPPL. FIG.}
% \renewcommand{\tablename}{SUPPL. TABLE}

\renewcommand{\theequation}{A\arabic{equation}}

\section{Neutrino energy fractions from \textsc{Pythia\,8}}
\label{app:pythia}

The neutrino energy fraction $\xi_\nu$ entering Eq.~(\ref{eq:Lnu_pp}) is conventionally taken as $\xi_\nu \simeq 0.05$ \cite{Kelner:2006tc}, being the fraction of the proton energy transferred to neutrinos of a single flavor in the leading-pion approximation. This estimate assumes single-pion production at threshold and neglects both the rising pion multiplicity with center-of-mass energy and the kaon contribution to the neutrino yield.

To obtain a more accurate determination, we perform \textsc{Pythia\,8.3}~\cite{Sjostrand:2014zea, Bierlich:2022pfr} simulations of inelastic $pp$ collisions at sixteen center-of-mass energies between $\sqrt{s} = 10$~GeV and $10^5$~GeV, generating $10^4$ minimum-bias events at each energy. For every event we collect all final-state muon- and electron-flavor neutrinos and antineutrinos, recording their energy fractions $x_\nu = E_\nu / E_p^{\rm lab}$ in the fixed-target frame. The per-flavor neutrino energy fraction is then computed as
\begin{equation}
\xi_\nu^{pp} (\sqrt{s}) = \left\langle \sum_{i \in \nu_\alpha} x_{\nu,i} \right\rangle_{\rm events},
\end{equation}
averaged over events and summed over all neutrinos of a given flavor in each event.

\begin{figure*}[t!]
\centering
\includegraphics[width=0.5\textwidth]{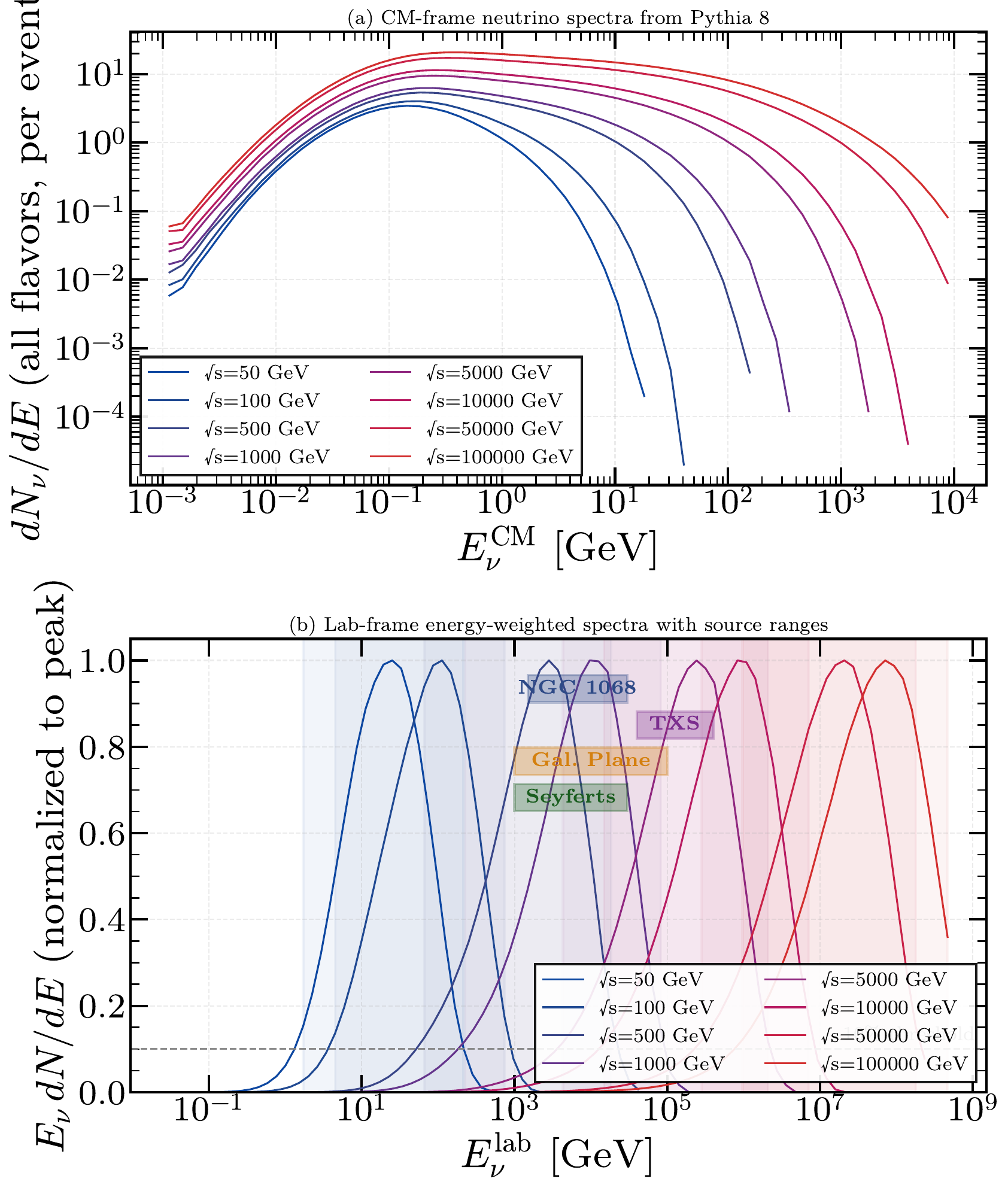}
\caption{\textbf{\textit{\textsc{Pythia\,8} neutrino energy distributions from inelastic $pp$ collisions.}} \textit{Top:} Center-of-mass frame spectra $dN_\nu/dE$ at selected $\sqrt{s}$ values. \textit{Bottom:} Lab-frame energy-weighted spectra $E_\nu\,dN/dE$, normalized to the peak, with vertical shaded regions showing the observed neutrino energy ranges for each source (NGC~1068, Galactic Center, TXS~0506+056, Seyfert stack). The horizontal dashed line indicates the 10\% threshold of the peak value used to define the effective energy range for each $\sqrt{s}$. }
\label{fig:pythia_dist}
\end{figure*}

\begin{figure*}[t]
\centering
\includegraphics[width=0.95\textwidth]{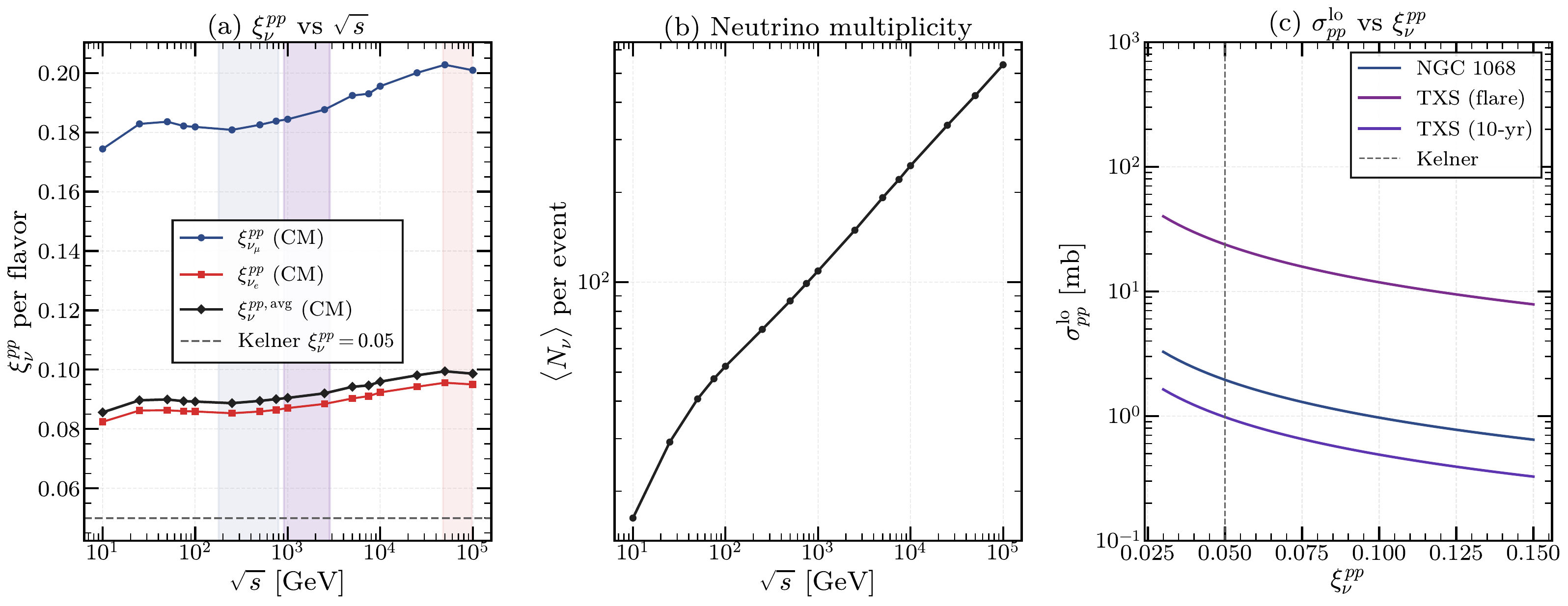}
\caption{\textbf{\textit{\textsc{Pythia\,8} neutrino properties from inelastic $pp$ collisions.}} \textit{Left:} Energy fraction $\xi_\nu^{pp}$ per flavor as a function of $\sqrt{s}$, with the Kelner {\it et al.} analytical estimate ($\xi_\nu^{pp} = 0.05$~\cite{Kelner:2006tc}) shown as a dashed line. Shaded vertical bands indicate the $\sqrt{s}$ ranges probed by each source (NGC~1068, Galactic Center, TXS~0506+056, Seyfert stack). \textit{Center:} Mean neutrino multiplicity per event. \textit{Right:} Lower bound on $\sigma_{pp}$ as a function of $\xi_\nu^{pp}$ for NGC~1068, TXS 0506, and the Galactic Plane,  illustrating the sensitivity of the cross-section constraint to the neutrino energy fraction.}
\label{fig:pythia_props}
\end{figure*}

The results are shown in Figs.~\ref{fig:pythia_dist} and~\ref{fig:pythia_props}. The upper panel of Fig.~\ref{fig:pythia_dist} displays the center-of-mass frame neutrino energy spectra $dN_\nu/dE$ at selected energies, while the lower panel shows the lab-frame energy-weighted spectra $E_\nu\,dN/dE$, normalized to the peak, which directly encode the contribution to $\xi_\nu^{pp}$. We find $\xi_\nu^{pp} \simeq 0.086$--$0.099$ per flavor across the relevant energy range, approximately a factor of $1.8$ larger than the canonical analytical value. This enhancement arises from the rising pion multiplicity and the increasing kaon fraction at higher energies.

The \textsc{Pythia}-derived $\xi_\nu^{pp} (\sqrt{s})$ is used self-consistently to determine the center-of-mass energies and cross-section bounds reported in the main text. For each source, the mapping $E_\nu \to \sqrt{s}$ is obtained by iterating $E_p = E_\nu / \xi_\nu^{pp} (\sqrt{s})$ and $\sqrt{s} = \sqrt{2\,m_p\,E_p + 2\,m_p^2}$ until convergence (typically within five iterations). For $p\gamma$ interactions, we perform analogous \textsc{Pythia\,8} simulations at seven center-of-mass energies from $\sqrt{s}=2.2$ to $5000$~GeV, finding $\xi_\nu^{p\gamma}\simeq 0.069$ near threshold, rising to $\simeq 0.097$--$0.108$ at $\sqrt{s}\gtrsim 50$~GeV. At the $\Delta(1232)$ resonance ($\sqrt{s}\simeq 1.232$~GeV) we retain the standard value $\xi_\nu^{p\gamma}=0.017$~\cite{Mucke:1998mk}. The energy-dependent $\xi_\nu^{p\gamma}(\sqrt{s})$ is used self-consistently in the $p\gamma$ bounds reported in the main text.

We note that the energy-distribution analysis above is focused on the $pp$ channel. For $p\gamma$ interactions the question arises of whether the neutrino spectral shape differs qualitatively due to the different cross-section energy dependence (the $\Delta(1232)$ resonance and its associated threshold structure, versus the smoothly rising $pp$ total cross section) and the different per-collision neutrino yields ($\xi_\nu^{p\gamma}\simeq 0.017$ at the $\Delta$ resonance, rising to $\simeq 0.07$--$0.11$ above, it says 0.097-0.108 at higher $\sqrt{s}$, versus $\xi_\nu^{pp} \simeq 0.09$). For a power-law CR spectrum $dN_{\rm CR}/dE\propto E^{-\Gamma}$ interacting with a quasi-monochromatic photon target (as is approximately the case for the $\sim 10$~keV X-ray fields considered here), the resulting neutrino spectrum above the $\Delta$ threshold also follows $\sim E_\nu^{-\Gamma}$: the spectral shape is dominated by the steepness of the CR flux, not by the energy dependence of $\sigma_{p\gamma}$. The main qualitative difference is a sharp low-energy cutoff in the $p\gamma$ neutrino spectrum set by the $\Delta$-production threshold $E_p\,\varepsilon_\gamma\gtrsim (m_\Delta^2-m_p^2)/2\simeq 0.15$~GeV$^2$, which is absent in $pp$. Above threshold, the spectral slopes and the resulting $\xi_\nu^{p\gamma}$ are similar in character to the $pp$ case. For these reasons, restricting the detailed energy-distribution discussion to the $pp$ channel is sufficient; the $p\gamma$ conclusions carry over with the caveat of the threshold shift.

Figure~\ref{fig:pythia_props} summarizes the key properties of the \textsc{Pythia} simulations. The left panel shows $\xi_\nu^{pp} (\sqrt{s})$ for muon and electron neutrino flavors separately, as well as the flavor-averaged value; the horizontal dashed line marks the Kelner et~al.~\cite{Kelner:2006tc} analytical estimate $\xi_\nu^{pp} = 0.05$. The shaded vertical bands indicate the $\sqrt{s}$ ranges probed by each neutrino source. The central panel displays the mean neutrino multiplicity $\langle N_\nu \rangle$ per event, which rises steeply with $\sqrt{s}$ and reaches $\sim 200$ at $\sqrt{s} = 10^5$~GeV. The right panel shows the impact of $\xi_\nu^{pp}$ on the lower bound on $\sigma_{pp}$: since $\sigma_{pp}^{\rm lo} \propto 1/\xi_\nu^{pp}$, the \textsc{Pythia} value $\xi_\nu^{pp} \simeq 0.09$ relaxes the lower bound by nearly a factor of~2 relative to the Kelner estimate, a non-negligible systematic shift.

\section{Dependence on the CR spectral index and neutrino spectral width}
\label{app:spectral}

The center-of-mass energy probed at each source depends on two independent effects: the spectral index $\Gamma$ of the CR injection spectrum, and the finite width of the neutrino energy distribution produced at each $\sqrt{s}$.

\begin{figure*}[t!]
\centering
\includegraphics[width=0.95\textwidth]{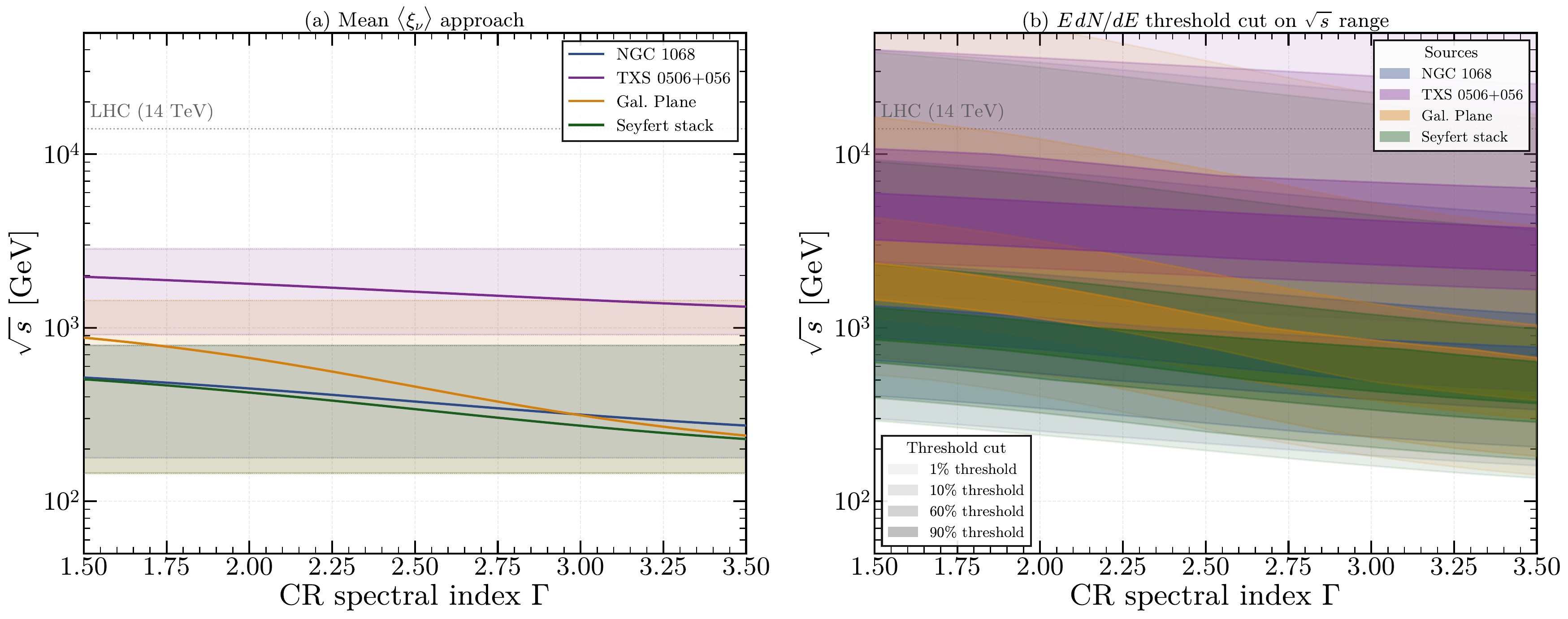}
\caption{\textbf{(a)} Effective center-of-mass energy $\sqrt{s}$ as a function of the CR spectral index $\Gamma$ for each neutrino source. Solid lines show the luminosity-weighted $\sqrt{s}$ computed from $\langle E_\nu \rangle_L$; shaded bands span the full observed $E_\nu$ range mapped to $\sqrt{s}$ via the \textsc{Pythia}-derived $\langle\xi_\nu\rangle$. Steeper spectra ($\Gamma \gtrsim 2.5$) weight lower neutrino energies, shifting $\sqrt{s}$ downward by up to a factor of $\sim 2$ relative to the hardest spectra ($\Gamma \simeq 1.5$). \textbf{(b)} Impact of threshold cuts on the $\sqrt{s}$ range. Nested bands show the $\sqrt{s}$ range for neutrinos with $E_\nu\,dN/dE_\nu$ above $1\%$, $10\%$, $60\%$, and $90\%$ of the spectral peak. Darker shading corresponds to tighter thresholds (neutrinos closer to the spectral peak). The horizontal dashed lines mark the LHC energy for reference.}
\label{fig:spectral_index}
\end{figure*}

The CR spectrum $dN_{\rm CR}/dE \propto E^{-\Gamma}$ weights the luminosity-averaged neutrino energy within the observed band through
\begin{equation}
\langle E_\nu \rangle_L = \frac{\int_{E_{\nu,\min}}^{E_{\nu,\max}} E_\nu^{2-\Gamma}\,dE_\nu}{\int_{E_{\nu,\min}}^{E_{\nu,\max}} E_\nu^{1-\Gamma}\,dE_\nu}\,,
\end{equation}
which shifts toward the low-energy end for steep spectra ($\Gamma \gg 2$) and toward the high-energy end for hard spectra ($\Gamma \lesssim 2$).
The resulting dependence of the effective $\sqrt{s}$ on $\Gamma$ is shown in Fig.~\ref{fig:spectral_index}a: varying $\Gamma$ from $1.5$ to $3.5$ shifts the effective center-of-mass energy by a factor of $\sim 2$ for each source. 
This weighting uses the intrinsic source spectrum; convolving with the IceCube effective area $A_{\rm eff}(E_\nu)\propto E_\nu^{0.5-1}$ would bias the detected weighted-mean $\sqrt{s}$ upward by a factor $\sim 2$--$3$ for steep spectra ($\Gamma\gtrsim 2.5$).

Additionally, the mean-$\xi$ prescription used in the main text assigns a single $\sqrt{s}$ to each observed $E_\nu$, but the lab-frame neutrino spectrum $E_\nu\,dN/dE_\nu$ produced at a given $\sqrt{s}$ has significant width (cf.\ Fig.~\ref{fig:pythia_dist}b), so an observed neutrino energy receives contributions from a range of parent $\sqrt{s}$ values.
To quantify this broadening, we define threshold-dependent energy fractions: for each simulated $\sqrt{s}$, we identify the neutrino energy boundaries where $E_\nu\,dN/dE_\nu \geq f\times\mathrm{peak}$ for $f = 1\%$, $10\%$, $60\%$, and $90\%$, and compute the corresponding $\xi_{\rm lo}(f)$ and $\xi_{\rm hi}(f)$. 
Using these in place of the mean $\langle\xi_\nu\rangle$ maps each observed $E_\nu$ to a $\sqrt{s}$ \emph{band} rather than a single value. 
Figure~\ref{fig:spectral_index}b shows the resulting bands for each source and threshold. 
Tighter cuts (e.g.\ $90\%$) isolate neutrinos near the spectral peak and give narrow $\sqrt{s}$ ranges close to the mean-$\xi$ value, while looser cuts (e.g.\ $1\%$) include the spectral tails and broaden the effective $\sqrt{s}$ range by up to an order of magnitude. 
The $10\%$ threshold, which captures the core of the distribution, broadens the $\sqrt{s}$ range by a factor of $\sim 2$--$3$.

Importantly, the cross-section bounds themselves are essentially independent of both effects, since they depend on $\xi_\nu$, which varies only weakly over the relevant $\sqrt{s}$ range.

\section{CR--CR collisions}
\label{app:crcr}

The bounds in the main text assume that one of the colliding particles is a target with low energy---a thermal proton or an ambient photon. One may also consider collisions between two relativistic CRs confined within the same acceleration region or propagating through the same environment.
For two CRs of energies $E_1$ and $E_2$, the center-of-mass energy is $\sqrt{s} \simeq \sqrt{2 E_1 E_2}$, vastly exceeding the $\sqrt{s} \simeq \sqrt{2 m_p E}$ available in CR--gas collisions at the same proton energy. For instance, two $100$~TeV CRs yield $\sqrt{s} \simeq 140$~TeV, an order of magnitude beyond the LHC.

The price of accessing such high center-of-mass energies is the steeply falling CR spectrum. Also, if the CRs are being accelerated radially outward (in jet-accelerated scenarios), the chance of CR-CR collision is very small. That is not necessarily the case for stochastic and turbulent acceleration scenarios, though. In CR--gas interactions, the target density is independent of energy, whereas in CR--CR collisions, the ``target'' flux falls as $E^{-\Gamma}$ with spectral index $\Gamma \simeq 2$--$3$. 
For the CR--CR channel to yield an interaction rate comparable to the CR--gas channel, the cross section must therefore grow with energy fast enough to compensate for the declining target flux. Concretely, if the cross section rises as a power law $\sigma_{pp} \propto (\sqrt{s})^p$ above the last energy at which the SM value is anchored, a spectral index $\Gamma$ requires $p \gtrsim 2(\Gamma - 1)$ to offset the falling CR density. 
For $\Gamma \simeq 1.75$--$2.5$ this corresponds to $p \simeq 1.5$--$3$.

\begin{figure}[t!]
\centering
\includegraphics[width=0.7\columnwidth]{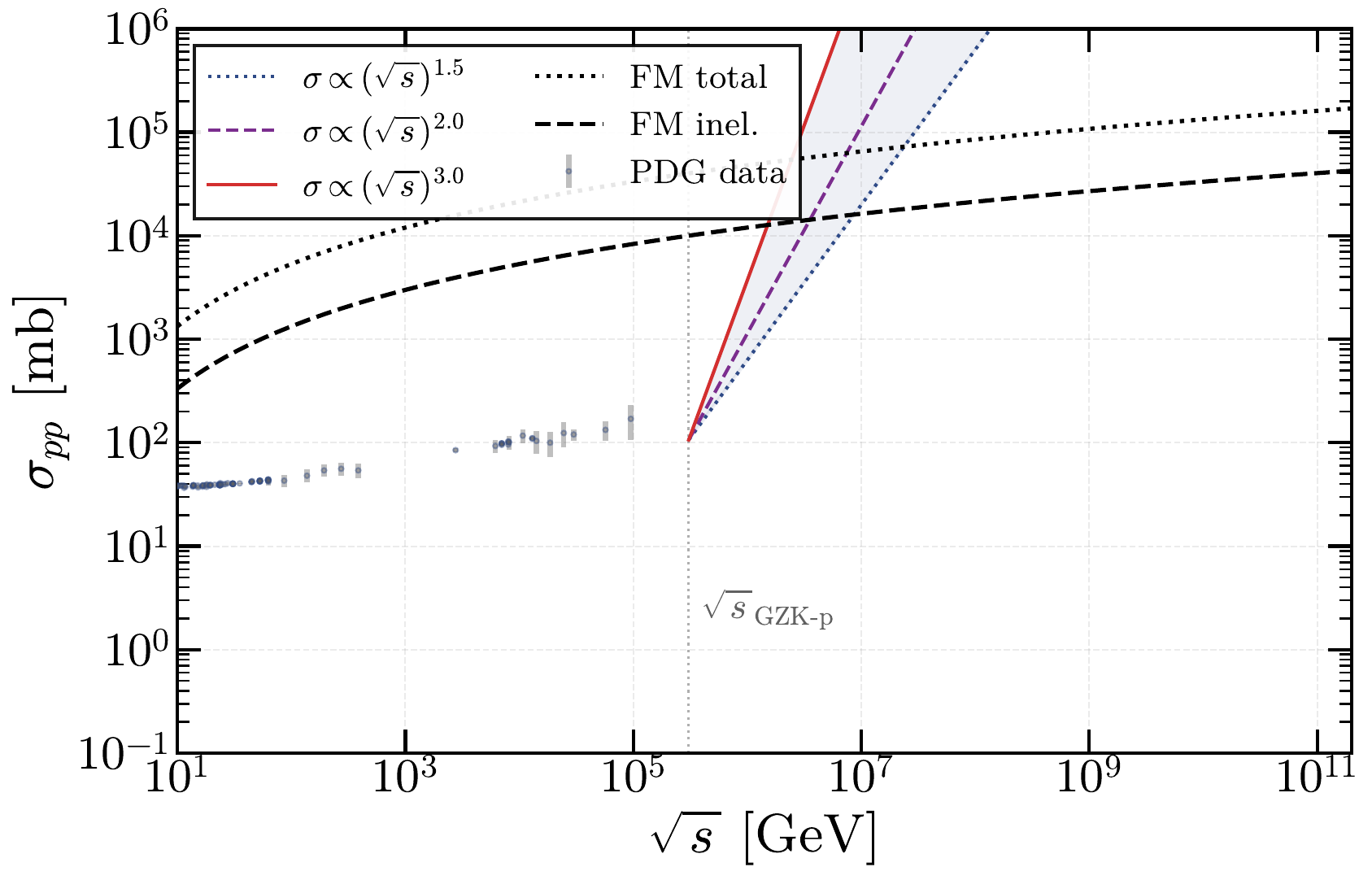}
\caption{\textbf{\textit{Power-law growths of the $pp$ cross section beyond $\sqrt{s}_{\rm GZK}$ (vertical dotted line), the center-of-mass energy of a GZK proton on a target proton at rest.}} The shaded band shows $\sigma \propto (\sqrt{s})^p$ with $p = 1.5$--$3$, the range needed to compensate the falling CR spectrum ($\Gamma \simeq 1.75$--$2.5$) and make CR--CR collisions competitive with CR--gas interactions. The dashed purple curve shows low-scale quantum gravity growth $\sigma \propto (\sqrt{s})^2$~\cite{Ettengruber:2025kat}. All extrapolations are anchored at the SM prediction and compared with the Froissart--Martin unitarity bounds (dashed/dotted black lines). Data points show PDG $pp$ measurements.}
\label{fig:limits_CRCR}
\end{figure}

In Fig.~\ref{fig:limits_CRCR} we illustrate this by anchoring the cross section at the value predicted by the SM at $\sqrt{s}_{\rm GZK}$---the center-of-mass energy of a GZK-scale proton ($E \sim 10^{20}$~eV) colliding with a proton at rest---and extrapolating with power laws $\sigma \propto (\sqrt{s})^p$ for $p \in [1.5, 3]$ up to the kinematic limit $\sqrt{s}_{\max} \simeq 2 E_{\rm GZK}$ for head-on CR--CR encounters.
For comparison, we also show the growth expected in low-scale quantum gravity scenarios where black-hole formation leads to $\sigma \propto (\sqrt{s})^2$~\cite{Ettengruber:2025kat}. All power-law extrapolations eventually exceed the Froissart--Martin unitarity bound~\cite{Froissart:1961ux, Martin:1965jj}---as expected, since any $\sigma \propto (\sqrt{s})^p$ growth with $p>0$ outpaces the $\ln^2(s)$ limit---indicating that such scenarios are unphysical at sufficiently high energies.

\end{document}